\documentclass[namedreferences]{solarphysics}
\usepackage[optionalrh,solaenum]{spr-sola-addons} 
\usepackage{graphicx}                    
\usepackage{color}                       
\usepackage{url}                         
\usepackage{multirow}
\usepackage[pdfborder={0 0 0 },urlcolor=blue,breaklinks]{hyperref}

\ifx \doiurl \undefined \def \doiurl#1{\href{http://dx.doi.org/#1}{\url{#1}}}\fi
\ifx \adsurl \undefined \def \adsurl#1{\href{http://adsabs.harvard.edu/abs/#1}{\url{#1}}}\fi

\newcommand{\etal}{{\it et al.}}

\newcommand{\lyra}{LYRA} 
\newcommand{\swap}{SWAP}
\newcommand{\proba}{PRO\-BA2} 
\newcommand{\goes}{GOES} 
\newcommand{\eve}{SDO/\-EVE} 
\newcommand{\picard}{\textit{Picard}}
\newcommand{\premos}{PRE\-MOS} 
\newcommand{\solstice}{SOR\-CE/\-SOLS\-TICE}
\newcommand{\see}{TIM\-ED/SEE}
\newcommand{\sem}{SOHO/SEM}
\newcommand{\xrs}{GOES/XRS}
\newcommand{\euvs}{GOES/EUVS}

\newcommand{\adv}{    {\it Adv. Space Res.}} 
 
\newcommand{\aap}{    {\it Astron. Astrophys.}}

\newcommand{\apj}{    {\it Astrophys. J.}}
\newcommand{\apjl}{   {\it Astrophys. J. Lett.}}

\newcommand{\grl}{    {\it Geophys. Res. Lett.}}

\newcommand{\jgr}{    {\it J. Geophys. Res.}}

\newcommand{\solphys}{{\it Solar Phys.}}

\usepackage{ifthen}
\usepackage{hyperref}
\newcommand{\solareceiveddatestring}[0]{
	\ifthenelse{\isundefined{\solareceiveddate}}{
    	 \textbullet\textbullet \textbullet\textbullet\textbullet\textbullet\textbullet \textbullet\textbullet\textbullet\textbullet
	}{
	\solareceiveddate{}
	}
}
\newcommand{\solaaccepteddatestring}[0]{
	\ifthenelse{\isundefined{\solaaccepteddate}}{
    	\textbullet\textbullet \textbullet\textbullet\textbullet\textbullet\textbullet \textbullet\textbullet\textbullet\textbullet
	}{
    	\solaaccepteddate{}
	}
}
\newcommand{\solaonlinedatestring}[0]{
	\ifthenelse{\isundefined{\solaonlinedate}}{
    		\textbullet\textbullet \textbullet\textbullet\textbullet\textbullet\textbullet \textbullet\textbullet\textbullet\textbullet
	}{
	\solaonlinedate{}
	}
}

\def\solavolume{\textbullet\textbullet\textbullet}
\def\solayear{\textbullet\textbullet\textbullet\textbullet}
\def\firstpage{0}
\def\lastpage{0}
\date{Received:\solareceiveddatestring{}/ Accepted:\solaaccepteddatestring{}/ Published online:\solaonlinedatestring{}}

\def\ps@opening{	
  \ifnum \lastpage =0 \def\lastpage{\textbullet\textbullet\textbullet} \else \fi
  \ifnum \firstpage=0 \def\solapages{\textbullet\textbullet\textbullet\,--\,\textbullet\textbullet\textbullet} \else \def\solapages{\firstpage\,--\,\lastpage}\fi
  \ifthenelse{\isundefined{\solareceiveddate}}{
    \def\@oddhead{\parbox[t]{\textwidth}{\footnotesize{}In preparation for submission to \textit{Solar Physics}\\
\url{http://www.springerlink.com/content/0038-0938}}}
	}{
    \def\@oddhead{\parbox[t]{\textwidth}{\footnotesize{}Submitted to \textit{Solar Physics},
\url{http://www.springerlink.com/content/0038-0938}}}
  }
  \ifthenelse{\isundefined{\solaaccepteddate}}{}
   {
    \def\@oddhead{\parbox[t]{\textwidth}{\footnotesize{}Accepted for publication in \textit{Solar Physics}, waiting for the authoritative version and a DOI,\\which will be available at
\url{http://www.springerlink.com/content/0038-0938}}}
   }
  \ifthenelse{\isundefined{\Xdoi}}{s11207-013-0252-5}
   {
    \def\@oddhead{\parbox[t]{\textwidth}{\footnotesize{}This is an emulation of \textit{Solar Physics} \textbf{\solavolume} \solapages, \solayear\\The authoritative version is DOI: \href{http://dx.doi.org/doi:10.1007/\Xdoi}{10.1007/\Xdoi}}}
   }
  \let\@evenhead\@oddhead
  \def\@oddfoot{\@gobble\idline\hfill}
  \def\@evenfoot{\hfill\@gobble\idline}
}
\def\copyright@text{}	
\def\@oddhead{\rh@rule\parbox[t]{\textwidth}{\@runningtitle\hfill\footnotesize{\thepage}}}
\def\@evenhead{\rh@rule\parbox[t]{\textwidth}{\footnotesize{\thepage}\hfill\@runningauthor}}

\newcounter{mspage}
\begin{document}

\begin{article}

\begin{opening}

\title{The LYRA Instrument Onboard PROBA2: Description and In-Flight Performance}

%
\author{M.~\surname{Dominique}$^{1}$\sep
        J.-F.~\surname{Hochedez}$^{1,2}$\sep
        W.~\surname{Schmutz}$^{3}$\sep
        I.E.~\surname{Dammasch}$^{1}$\sep
        A.I.~\surname{Shapiro}$^{3}$\sep
        M.~\surname{Kretzschmar}$^{1,4}$\sep
        A.N.~\surname{Zhukov}$^{1,5}$\sep
       D.~\surname{Gillotay}$^{6}$\sep
       Y.~\surname{Stockman}$^{7}$\sep
       A.~\surname{BenMoussa}$^{1}$
      }

%
\runningauthor{M. Dominique \etal}
\runningtitle{LYRA Instrument}


\institute{$^{1}$ Solar-Terrestrial Center of Excellence - SIDC, Royal Observatory of Belgium, Brussels, Belgium
              email: \url{marie.dominique@oma.be}\\ 
             $^{2}$ LATMOS, Guyancourt, France\\
             $^{3}$ PMOD/WRC, Davos, Switzerland\\
             $^{4}$ LPC2E, Universit\'e d'Orl\'eans, Orl\'eans, France\\ 
             $^{5}$ Skobeltsyn Institute of Nuclear Physics, Moscow State University, Moscow, Russia\\
             $^{6}$ Belgian Institute for Space Aeronomy, Brussels, Belgium\\
             $^{7}$ Centre Spatial de Li\`ege, Li\`ege, Belgium\\            
             }

\begin{abstract}
The \textit{Large Yield Radiometer} (\lyra) is an XUV--EUV--MUV (soft X-ray to mid-ultraviolet) solar radiometer onboard the European Space Agency  \proba \- mission that was launched in November 2009.  \lyra \- acquires solar irradiance measurements at a high cadence (nominally 20\,Hz) in four broad spectral channels, from soft X-ray to MUV, that have been chosen for their relevance to solar physics, space weather and aeronomy. In this article, we briefly review the design of the instrument, give an overview of the data products distributed through the instrument website, and describe the way that data are calibrated. We also briefly present a summary of the main fields of research currently under investigation by the \lyra\-consortium.
\end{abstract}

%

\end{opening}

%
 \section{Introduction}\label{s:intro}

The \textit{Large Yield Radiometer} (\lyra: \opencite{2006AdSpR..37..303H}), is an XUV to MUV (soft X-ray to mid=ultraviolet) solar radiometer embarked on the European Space Agency \proba \- mission that was launched on 2 November 2009. It revolves on a Sun-synchronous orbit with an altitude around 720\,km, and faces the Sun continuously.  \lyra \- acquires solar-irradiance measurements in four broad spectral channels, from soft X-ray to UV, that have been chosen for their relevance to solar physics, space weather and aeronomy. 

The objective of the instrument is twofold: 
\begin{itemize}
\item providing time series of solar irradiance with a very high sampling cadence (up to 100\,Hz) in spectral ranges complementary to other active radiometers, \textit{e.g.} \textit{Geostationary Operations Environmental Satellite/Extreme Ultraviolet Sensor} (\euvs), \textit{Solar and Heliospheric Observatory/Solar Extreme Ultraviolet Monitor} (\sem: \opencite{1998SoPh..177..161J}), \textit{Solar Radiation and Climate Experiment/Solar Stellar Irradiance Comparison Experiment}(\solstice: \opencite{1993JGR....9810667R}; \opencite{1993JGR....9810679W}), \textit{Thermosphere Ionosphere Mesosphere Energetics and Dynamics/Solar EUV Experiment} (\see: \opencite{2004SPIE.5660...36W}), \picard/\textit{Precision Monitor Sensor} (\premos: \opencite{2009Metro..46S.202S}), \textit{Solar Dynamics Observatory/EUV Variability Experiment} (\eve: \opencite{springerlink:10.1007/s11207-009-9487-6}). 
\item testing an innovative kind of wide-bandgap diamond detectors. Such detectors, being radiation-hard and visible-blind, are particularly well adapted to observe the wavelengths targeted by \lyra. 
\end{itemize}

The instrument and its pre-flight calibration were described by \inlinecite{2009A&A...508.1085B}. Here, we review the design of the instrument in Section \ref{s:design}. Section \ref{s:data}  provides an overview of the data products that are distributed to the scientific community and details their calibration. Section \ref{s:sc_opportunities} briefly presents the main fields of scientific research currently investigated, exploiting \lyra \- data.

\begin{figure}[h!]    
   \centerline{\includegraphics[width=0.45\textwidth,angle=0, clip=, trim=0 0 70 0]{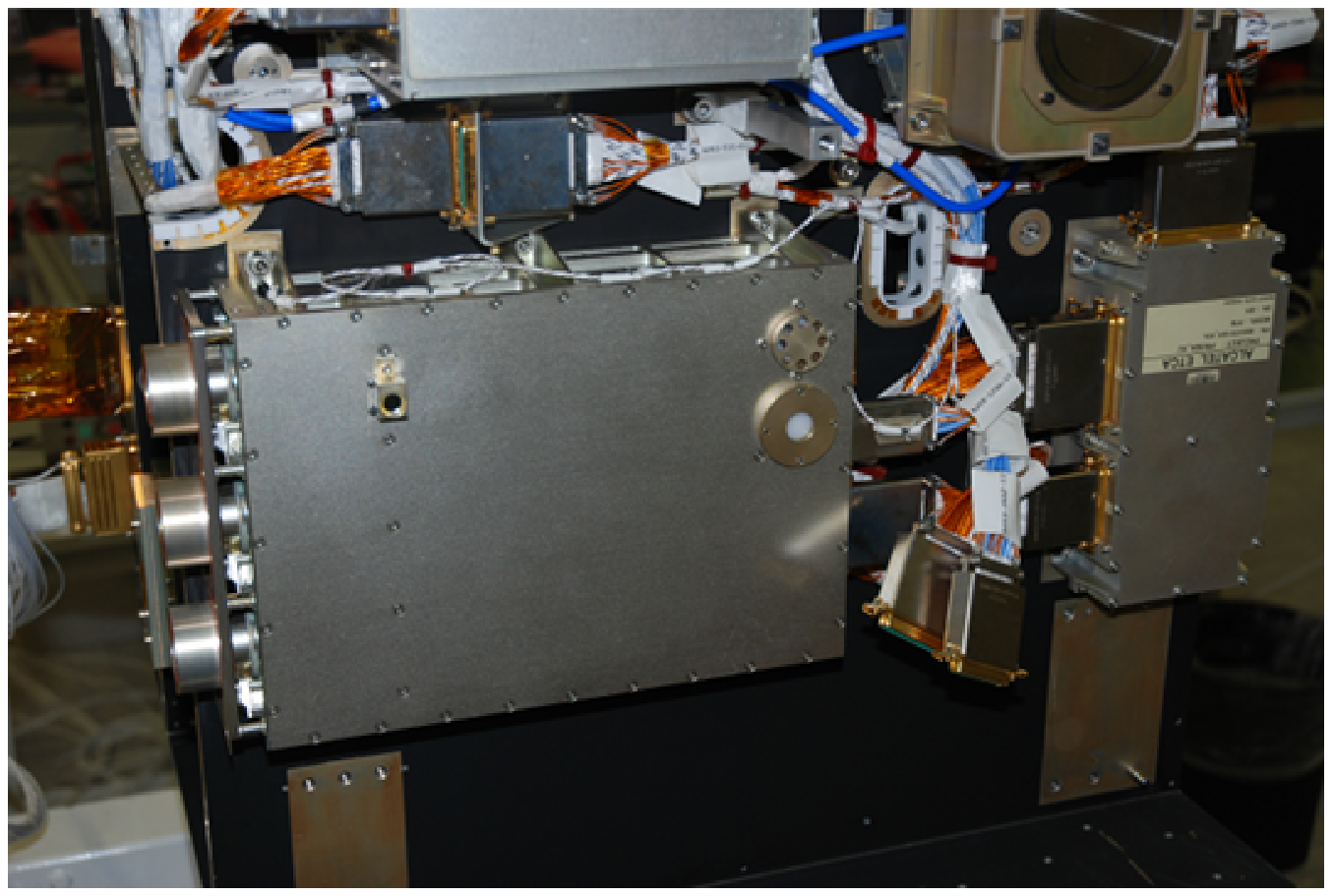} \hspace{0.1\textwidth}\includegraphics[width=0.24\textwidth,angle=0, clip=]{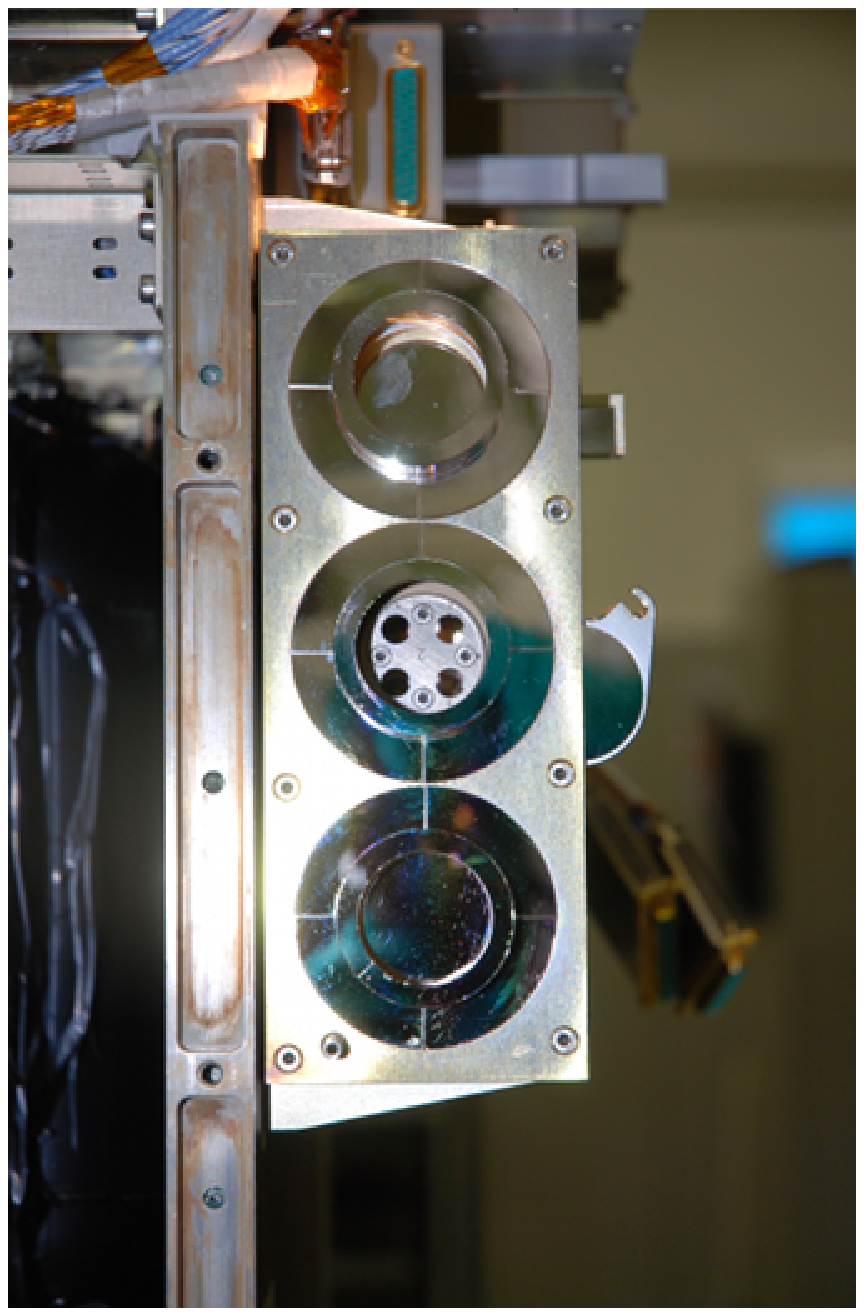}}
     \vspace{-4.5cm}   
     \centerline{ \bf     
      \hspace{0.1\textwidth}   \colorbox{black}{\color{white}{\bf{(a)}}}
      \hspace{0.48\textwidth}  \colorbox{black}{\color{white}{\bf{(b)}}}
         \hfill}
     \vspace{4.5cm}    
\caption{\lyra \- (315 $\times$ 92.5 $\times$ 222\,mm$^{3}$) during its integration on \proba. Panel (b), the open cover of unit 2 reveals the four observation channels.
        }
   \label{fig:lyra}
\end{figure}

\section{Design of the Instrument}\label{s:design}

\begin{figure}
   \centerline{\includegraphics[width=0.8\textwidth,clip=]{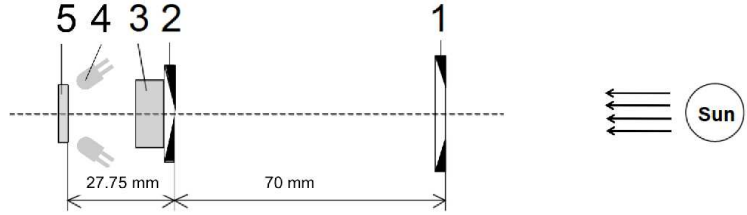}
              }
   \caption{Schematic representation of one \lyra \- channel: 1 and 2 are respectively the view-limiting and precision apertures, together forming the collimator, 3 is the filter, 4 are the two LEDs at 375 and 470 nm located behind the filter, and 5 is the detector.
               }
   \label{fig:optics}
\end{figure}

\lyra \- is a shoe-box size instrument (315 $\times$ 92.5 $\times$ 222\,mm$^3$) composed of three quasi-redundant units, each equipped with an individual cover and hosting four spectral channels (see Figure \ref{fig:lyra}). A channel consists of a collimator, an optical filter, a detector, and two LEDs on the side. LEDs are located between the filter and the detector and emit at 375 and 470 nm (see Figure \ref{fig:optics}). They are used to estimate the impact of ageing on the detectors. 
   
The three units are essentially similar in terms of spectral coverage, but involve non-identical associations of filters and detectors (Table \ref{tab:channels}). The four bandpasses are indicated in this table, as well as their purity (\textit{i.e.} the ratio of the flux in the defined bandwidth to the total output signal). Purities have been computed considering a quiet-Sun-type spectrum. They show that for all channels but Lyman-$\alpha$, most of the detected signal actually comes from the defined bandpasses. The Lyman-$\alpha$ channel, however, is highly contaminated by out-of-band radiation.    

\begin{table}
\caption{Summary of the characteristics of \lyra \- channels. The purity is defined as the ratio of the flux in the nominal wavelength range (defined bandwidth) to the total output signal. Purities correspond to a solar-minimum-type spectrum. Thicknesses of aluminum and zirconium filter layers are indicated in the ``filter label" column, together with the Acton reference for Lyman-$\alpha$ and Herzberg filters.}\label{tab:channels}
\begin{tabular}{ccccc}     
\hline                  
\textbf{Channel} & \textbf{Filter label} & \textbf{Detector} & \textbf{Bandwidth} & \textbf{Purity}\\
\hline
\textit{Unit 1} & & & & \\
1-1 & Lyman-$\alpha$ [122XN] & MSM Diamond & 120\,--\,123\,nm &  26\,\% \\
1-2 & Herzberg [220B] & PIN Diamond & 190\,--\,222\,nm &  95\,\% \\
1-3 & Aluminum (158 nm) & MSM Diamond & 17\,--\,80\,nm + $<$ 5\,nm & 96.8\,\% \\
1-4 & Zirconium (300 nm) &  AXUV Si & 6\,--\,20\,nm + $<$ 2\,nm & 97\,\% \\
\hline
\textit{Unit 2} & & & & \\
2-1 & Lyman-$\alpha$ [122XN] & MSM Diamond & 120\,--\,123\,nm &  25.7\,\% \\
2-2 & Herzberg [220B] & PIN Diamond & 190\,--\,222\,nm &  95\,\% \\
2-3 & Aluminum (158 nm) & MSM Diamond & 17\,--\,80\,nm + $<$ 5\,nm & 97.2\,\% \\
2-4 & Zirconium (141 nm) &  MSM Diamond & 6\,--\,20\,nm + $<$ 2\,nm &  92.2\,\% \\
\hline
\textit{Unit 3} & & & & \\
3-1 & Lyman-$\alpha$ [122N+XN] & AXUV Si & 120\,--\,123\,nm &   32.5 \,\%\\
3-2 & Herzberg [220B] & PIN Diamond & 190\,--\,222\,nm &  95\,\% \\
3-3 & Aluminum (158 nm) & AXUV Si & 17\,--\,80\,nm + $<$ 5\,nm & 96.6\,\% \\
3-4 & Zirconium (300 nm) &  AXUV Si & 6\,--\,20\,nm + $<$ 2\,nm & 95\,\% \\
\hline
\end{tabular}
\end{table}

\begin{figure}    
   \centerline{\hspace{0.25cm} \includegraphics[width=0.38\textwidth,angle=0, clip=, trim=0 2 0 2]{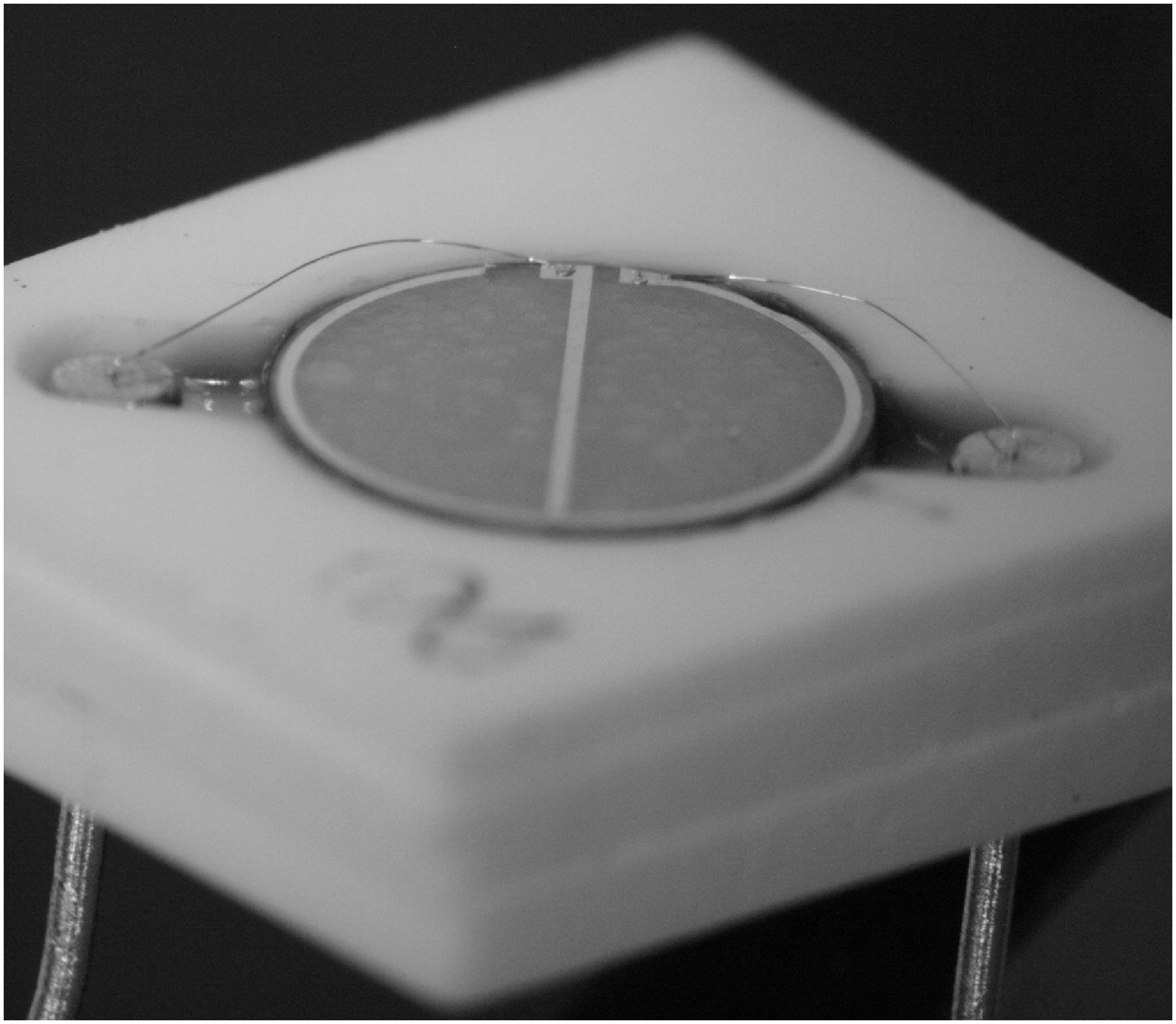} \hspace{0.12\textwidth}\includegraphics[width=0.38\textwidth,angle=0, clip=]{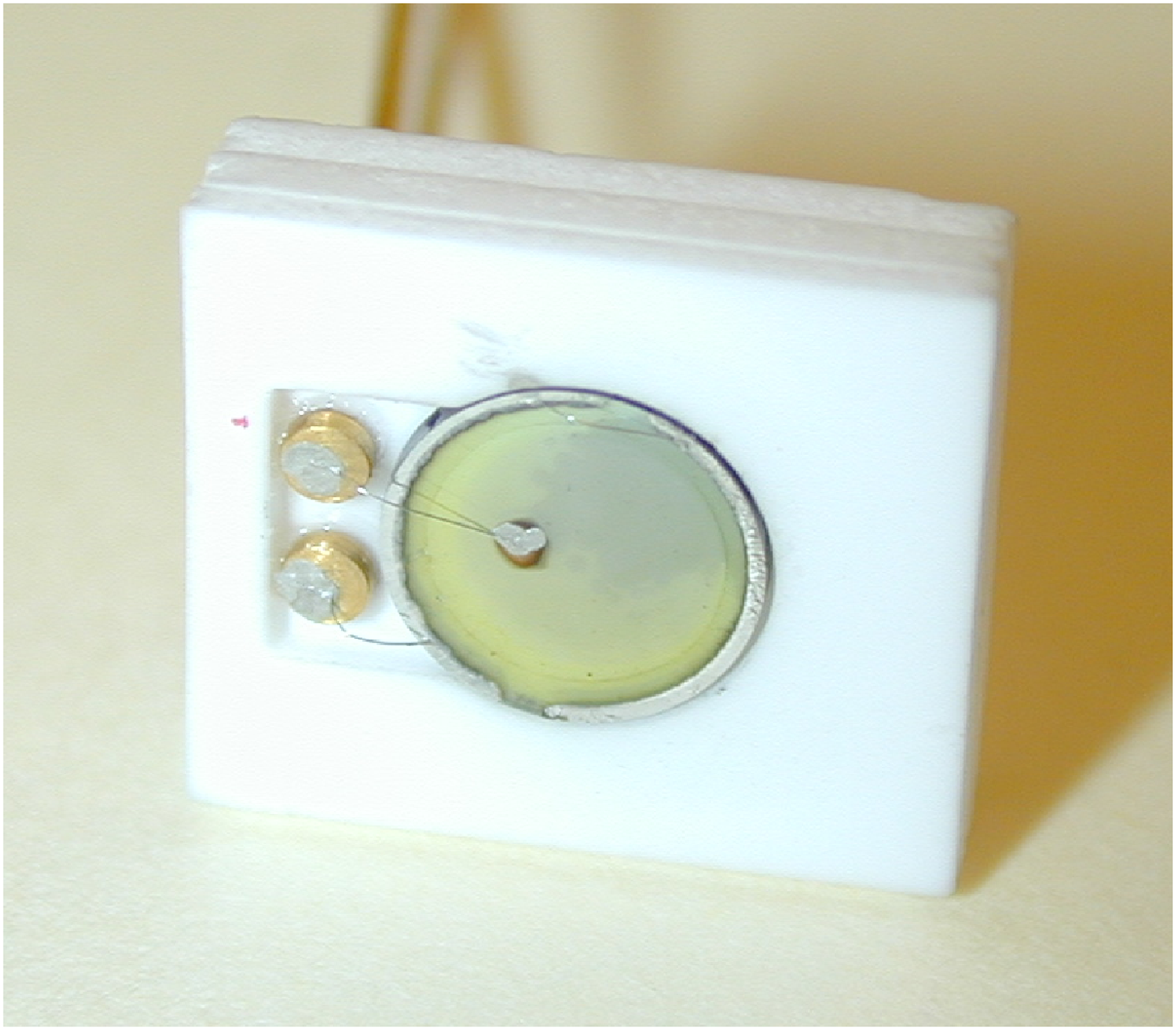}}
   \vspace{0.3cm}   
   \centerline{\includegraphics[width=0.49\textwidth,angle=0, clip=, trim=20 20 40 30]{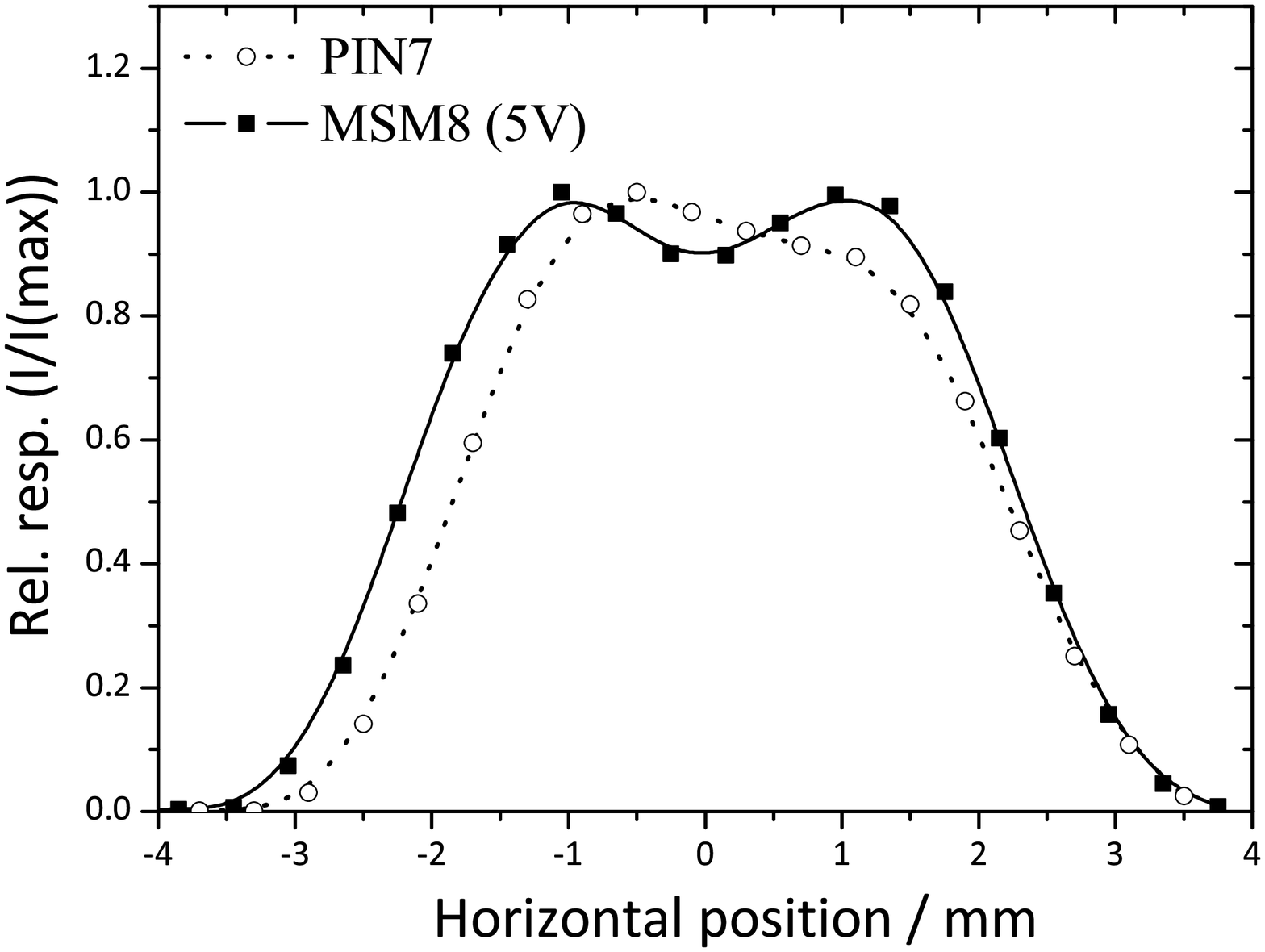} \hspace{0.01\textwidth}\includegraphics[width=0.49\textwidth,angle=0, clip=, trim=20 20 40 30]{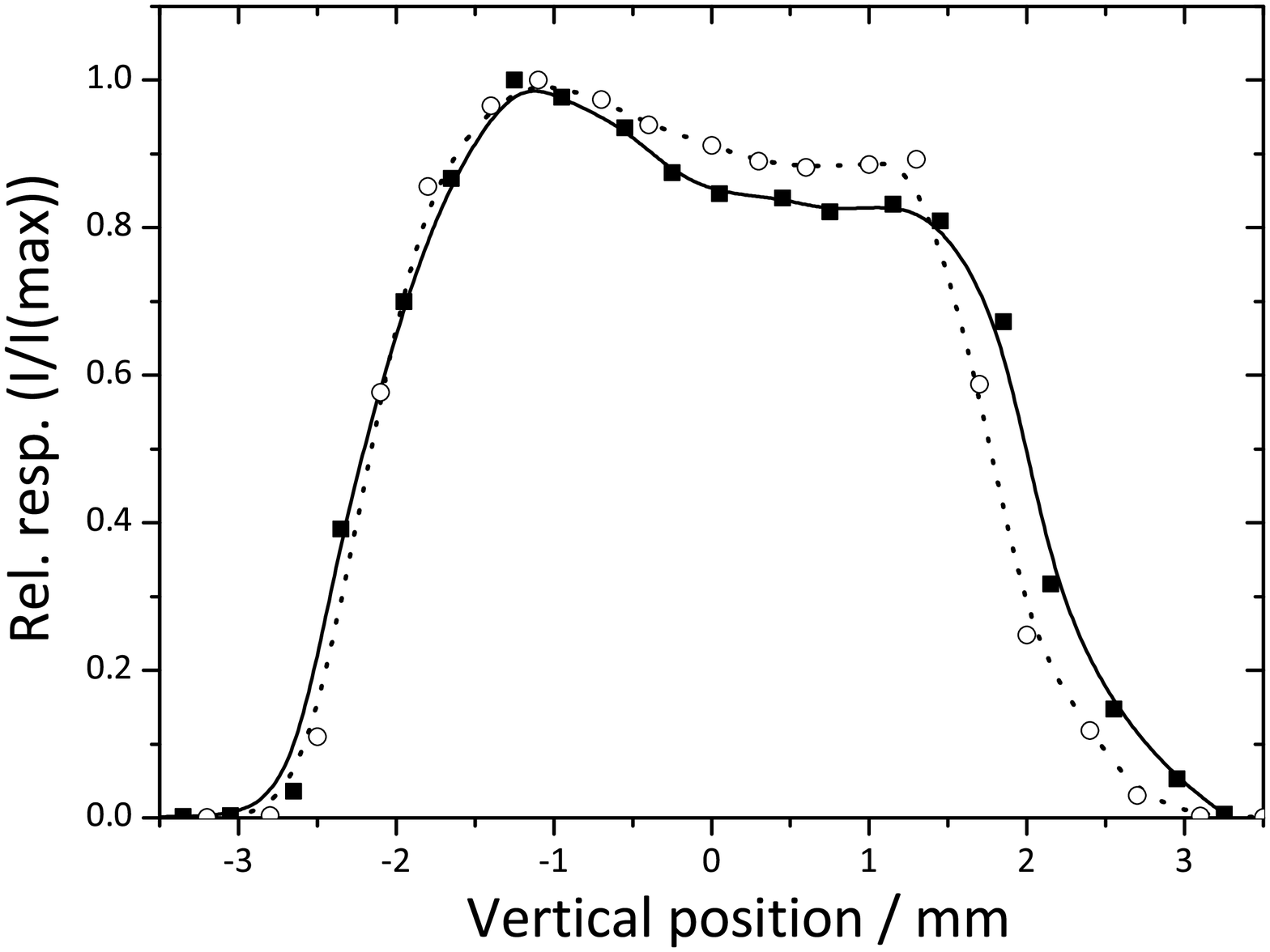}}
     \vspace{-8.5cm}   
     \centerline{ \bf     
      \hspace{0.36\textwidth}   \color{white}{(a)}
      \hspace{0.45\textwidth}  \color{black}{(b)}
         \hfill }
     \vspace{4.2cm}   
     \centerline{\bf     
      \hspace{0.36\textwidth}   \color{black}{(c)}
      \hspace{0.45\textwidth}  \color{black}{(d)}
         \hfill}
     \vspace{-3.cm}    
     \centerline{\hspace{1.7cm} \color{black}{\textbf{$<$---------------$>$}}
        \hfill }
     \vspace{0.0cm}    
     \centerline{\textbf{\hspace{2.5cm}   \color{black}{5\,mm}}  
        \hfill }
     \vspace{-0.2cm}
     \centerline{\hspace{8.35cm} \color{black}{\textbf{$<$---------$>$}} 
     	\hfill}
     \vspace{0.0cm}    
     \centerline{\textbf{\hspace{8.7cm}   \color{black}{5\,mm}}
      	\hfill}
     \vspace{5.5cm}    
\caption{Two examples of diamond detectors: MSM (a) and PIN (b). In both cases, we note the position of an electrode (central bar for the MSM, off-center circle for the PIN). Those electrodes strongly affect the flat-field of the detectors, as shown in the plots of relative responsivity along two perpendicular directions in (c) and (d).
        }
   \label{fig:detectors}
\end{figure}

\begin{figure}    
   \centerline{\includegraphics[width=1.0\textwidth,angle=0, clip=, trim=0 0 0 0]{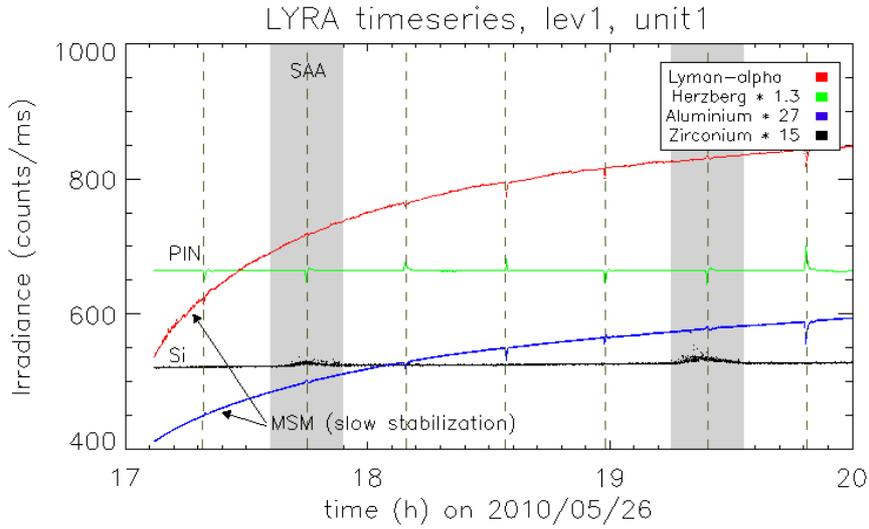}
   }
\caption{Slow stabilization of MSM detectors: the signal has not yet reached its stabilization level three hours after the detector was switched on. Also illustrated on this figure is the perturbation in the Si-detectors' signal when crossing the SAA. Diamond technology has proven to be more robust than silicon to the impact of high energy protons (and their secondary electrons) that causes this noisy behavior. Gray bands indicate the SAA and dashed vertical lines the large-angle rotations of the spacecraft. Time series have been rescaled to fit the same range and appear in the same order as in the legend. Scaling coefficients are indicated in the legend.} 
   \label{fig:SAA}
\end{figure}

In addition to wide-bandgap diamond detectors prototypes, units 1 and 3 incorporate some classical silicon (Si) detectors to allow the comparison between both technologies. Diamond  detectors are of two types: metal--semiconductor--metal (MSM) photoconductors and positive--intrinsic--negative (PIN) photodiodes (Figure \ref{fig:detectors}), the latter being used for the Herzberg channels only (see Table \ref{tab:channels}). Characteristics of these detectors are detailed by \inlinecite{2004PSSAR.201.2536B} and \inlinecite{2006NIMPA.568..398B}. A few of these characteristics are reminded here, since they significantly affect \lyra \- data: 
\begin{itemize}
\item An electrode is located on the surface of both MSM and PIN detectors, which results in important flat-field variations (see Figure \ref{fig:detectors}).
\item Trapping/detrapping of generated photoelectrons by defects (not bulk but surface defects) causes the signal of MSM detectors to take quite a long time before reaching stabilization (see Figure \ref{fig:SAA}).
\end{itemize}

\begin{figure}
   \centerline{\includegraphics[width=1\textwidth,clip=]{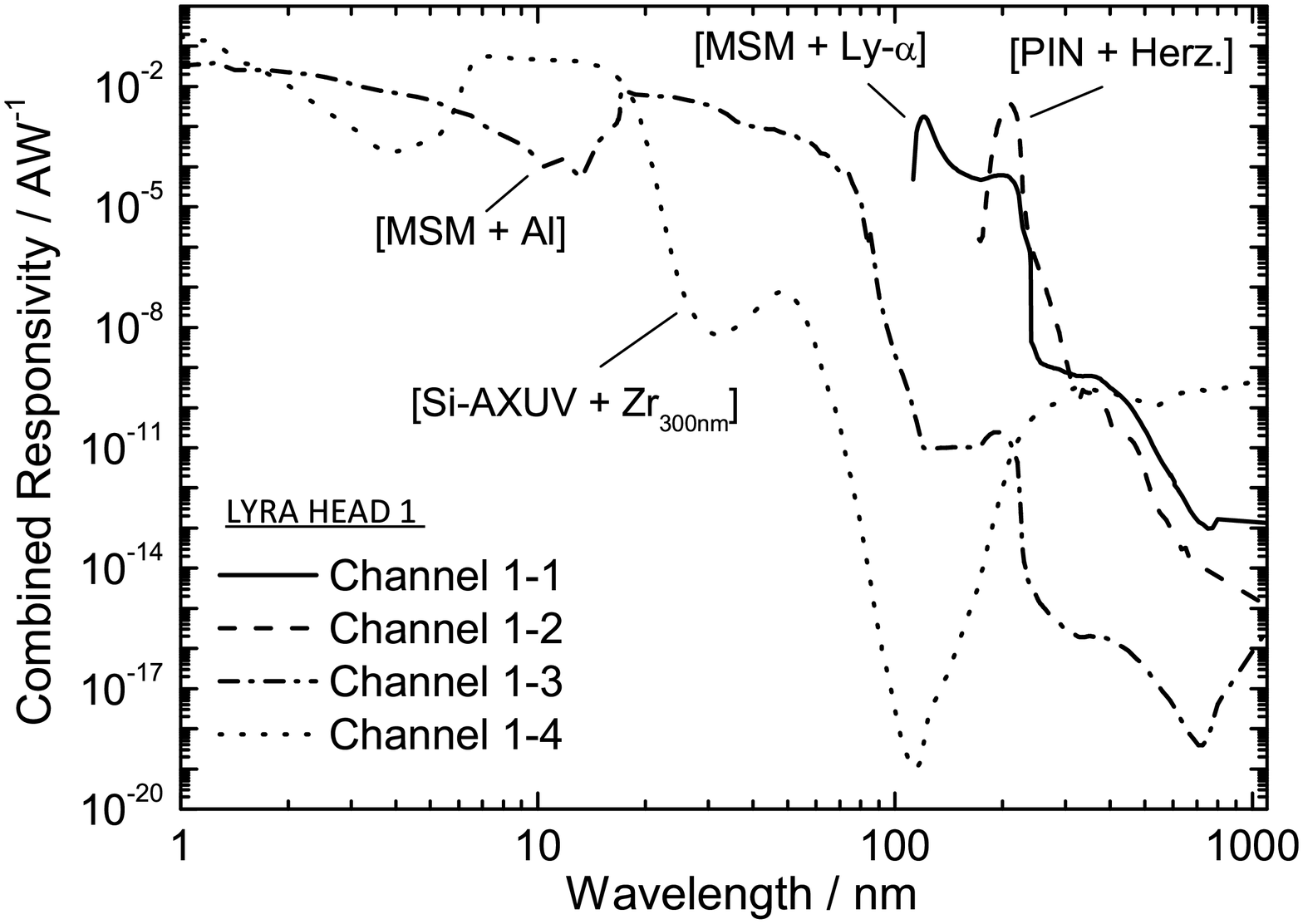}
              }
   \caption{Simulated combined spectral (detector + filter) responsivities for \lyra \- unit 1 between 1 and 1000\,nm, from \protect \inlinecite{2009A&A...508.1085B}.}
   \label{fig:responsivity}
\end{figure}

Figure \ref{fig:responsivity} illustrates the total responsivity of filter-detector combinations for all four channels of unit 1. These curves result from a radiometric model which uses pre-launch measurements of filter transmittance and detector responsivity as input. 

Raw \lyra \- data, as transmitted by the spacecraft, result from an on-board treatment of photocurrents produced by the detectors. Photocurrents are first converted to voltages in the 0\,--\,5 volts range by a resistor, of which the value is different for each channel. Voltages are then switched to frequencies by a voltage-to-frequency converter (VFC). Regularly, those VFC are connected to stable calibrated onboard voltage sources of 5 volts, 2.5 volts, and 0 volts. These measurements are interleaved in the telemetry, to be used as reference when processing the data. At the end of the acquisition chain there is a counter that counts the number of pulses received over the duration of the integration period. This number, which is expressed in counts, constitutes the raw \lyra \- data that are downloaded as part of the telemetry. 

\section{LYRA Data Description}\label{s:data}
\subsection{Data Products}
\lyra \- produces time series of spectral irradiance in its four bandpasses in a quasi-uninterrupted way. Interruptions might happen during calibration campaigns (once every two weeks on average), or during the Winter occultation season, when the spacecraft transits the Earth's shadow. Nevertheless, since \proba \- is flying on a polar, dawn--dusk heliosynchronous orbit, such occultations are limited to a three--four months winter period (depending on the channel) and last for maximum 25 minutes every orbit (one orbit is 100 minutes long).

As mentioned in Section \ref{s:design}, \lyra \- includes three units, which are similar from the spectral point of view. The four channels of a unit are operated in parallel, acquiring irradiance measurements at a nominal cadence of 20 Herz, but that could go up to 100 Herz.

The strategy behind the redundancy is the following:
\begin{itemize}
\item The nominal unit (unit 2) is in permanent use, but it is therefore the most affected by degradation. In the Lyman-$\alpha$ and Herzberg channels of this unit, the signal has dropped by about 99\,\% since the beginning of the mission. The degradation is so strong that the solar signal is now barely detectable in those two channels.
\item Unit 3 is used in a campaign-driven way and keeps its cover closed the rest of the time, limiting ageing effects.
\item Unit 1 is kept mostly unused, acquiring data for 40 minutes every three months on average and is therefore the most preserved. It is aimed at being a reference to estimate the degradation of units 2 and 3.
\end{itemize}

Data are usually available within four hours following their acquisition. Products with different levels of processing are distributed through the \proba \- website (\url{http://proba2.oma.be}). They consist of uncalibrated and calibrated data, as well as various quicklook datasets (see Table \ref{tab:lyradata}). 

Once downloaded, raw \lyra \- data in counts are converted back to 
\newline counts\,ms$^{-1}$, dividing them by the integration time, and distributed without further processing through the instrument website as \textbf{level-1/engineering data}, in standard files -- either from the nominal unit (\textsf{.std}) or from an additional `back-up" unit (\textsf{.bst}), that can be either unit 1 or 3. Information about the acquisition context (temperature, pointing, status of covers, status of LEDs, usage of a back-up unit $\ldots$) compose the ancillaries stored in the metadata (\textsf{.met}) file. Additional files gather the data acquired during calibration campaigns (dark currents, LED signals), or the rejected data (outliers, data acquired during transitions between acquisition modes).

\textbf{Level-2/calibrated data} are available through the same website. Currently, calibration includes subtraction of dark currents (which removes the temperature effects), compensation for degradation, rescaling to one astronomical unit, and conversion from counts\,ms${-1}$ to irradiance units, but no correction for flat-field effects yet. Special features, such as imprints of large angle rotations of the spacecraft or perturbations due to the South Atlantic Anomaly, are also visible in the data. Those features are detailed in Section \ref{s:features}. 

Basic IDL procedures to download and read \lyra \- data are available through the SolarSoft library (\url{sohowww.nascom.nasa.gov/solarsoft/}) at the address \newline \textsf{ssw/proba2/lyra/idl}.  

\begin{table}
\caption{ Summary of \lyra \- data products distributed to the scientific community. These products are processed after each data download (\textit{i.e.} every three to four hours).
}
\label{tab:lyradata}
\begin{tabular}{llll} 
  \hline
\textbf{Product} & \textbf{File extension on} & \textbf{Format} &  \textbf{Characteristics} \\
                         & \textbf{LYRA website} &                     &  \\
  \hline
  & 	\textsf{$\ast$\_lev1\_std(bst).fits}  & FITS  & unprocessed solar irradiance,  \\
  &&& in $\lbrack$counts\,ms$^{-1}\rbrack$ \\
Level 1  & \textsf{$\ast$\_lev1\_cal(bca).fits} & FITS & unprocessed calibration data, \\
engineering data  &&& in $\lbrack$counts\,ms$^{-1}\rbrack$  \\
 & \textsf{$\ast$\_lev1\_met.fits}  & FITS & ancillary data:   \\
 &&& temperature, pointing $\ldots$\\
  & \textsf{$\ast$\_lev1\_rej(bre).fits} & FITS & rejected samples (outliers $\ldots$) \\
 \hline
Level 2  & 	\multirow{2}{*}{\textsf{$\ast$\_lev2\_std.fits}} & \multirow{2}{*}{FITS} & calibrated solar irradiance, \\
basic science data &&& in $\lbrack$Wm$^{-2}\rbrack$ \\
  \hline
Level 3  & \multirow{2}{*}{\textsf{$\ast$\_lev3\_std.fits}}  & \multirow{2}{*}{FITS} & \multirow{1}{*}{level 2 averaged over 1 min,}\\
 averaged science data &&& in $\lbrack$Wm$^{-2}\rbrack$\\
  \hline
Level 4 A & \multirow{2}{*}{\textsf{$\ast$.png}}  & \multirow{2}{*}{image} &  \multirow{1}{*}{daily plot of calibrated data}\\
quicklooks  &  &  &  for all \lyra \- channels\\
  \hline
Level 4 B & \multirow{3}{*}{\textsf{$\ast$.png}} & \multirow{3}{*}{image}  &  \multirow{1}{*}{3-days \goes-like plot of }\\
 \multirow{2}{*}{quicklooks} & & &  calibrated data in aluminum \\
 &&& and zirconium channels \\
   \hline
Level 5 & \multirow{2}{*}{\textsf{html}} & \multirow{2}{*}{text file}  & \multirow{1}{*}{List of flares with links to}\\
flare list &&& \lyra \- and \goes \- flux profiles\\
   \hline
\end{tabular}
\end{table}

\subsection{Calibration of \lyra \- Data}\label{s:calibration}

\subsubsection{Dark Current Subtraction}
The dark currents were measured as a function of temperature between $-$40$^{\circ}$C and +60$^{\circ}$C in the laboratory before the launch, but only in steps of 10$^{\circ}$C. The relationship between temperature and dark current was globally linear below 40$^{\circ}$C (\textit{i.e.} over the estimated operational temperature range), which explains why no further test was performed with smaller steps. Unfortunately, it turned out that the onboard temperature experienced in space was much hotter than expected -- actually between +35$^{\circ}$C and +55$^{\circ}$C -- in a range where the functional relationship between temperature and dark current is non-linear. 

\begin{figure}      
    \centerline{\includegraphics[width=0.79\textwidth,clip=,
                angle=0]{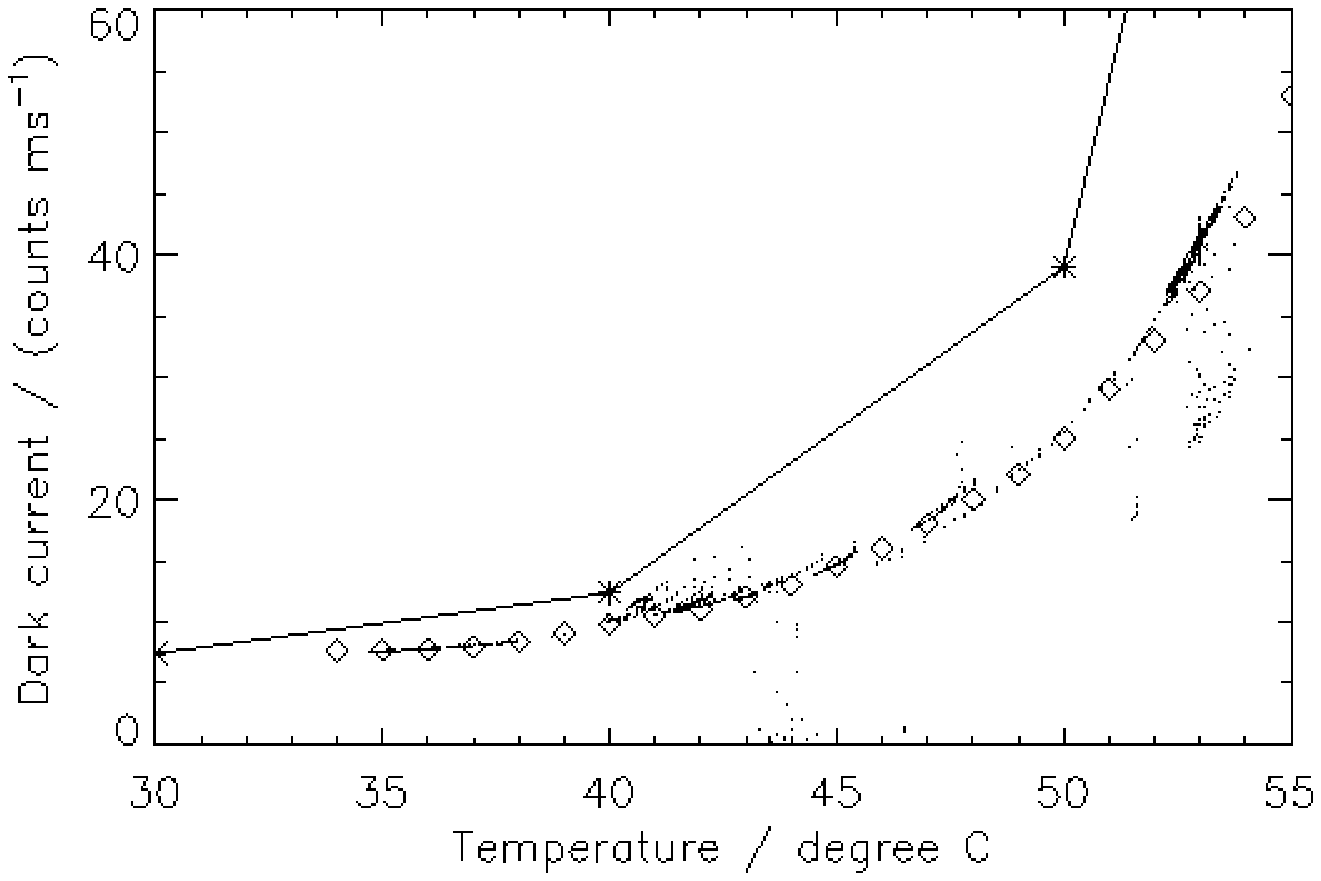}}
    \centerline{\hspace{-0.4cm}\includegraphics[width=0.83\textwidth,clip=,
                angle=0]{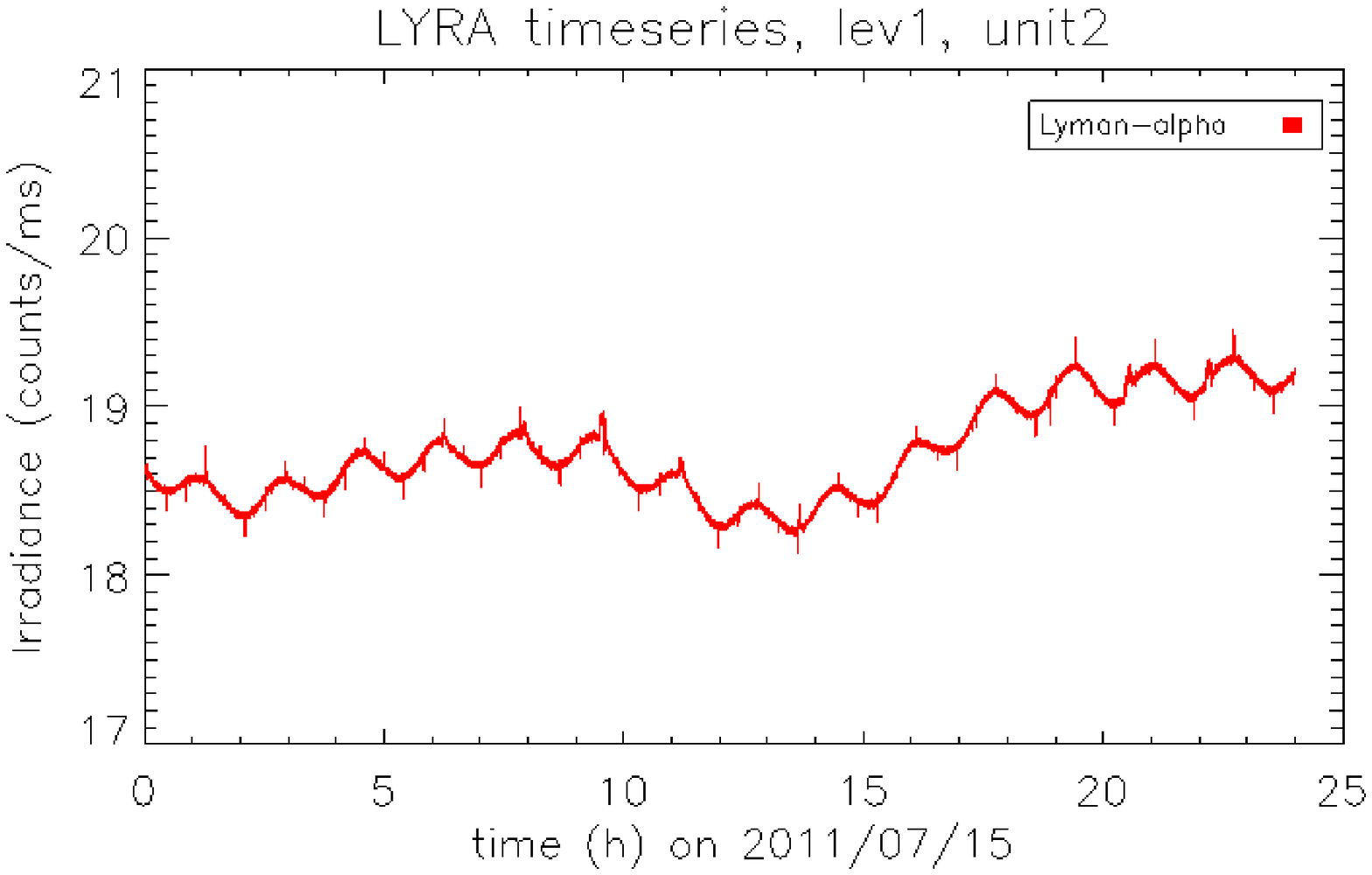} 
              }
     \vspace{-11.5cm}   
    \centerline{\hspace{-5cm}   \color{black}{(a)}}
     \vspace{6cm}   
     \centerline{\hspace{-5cm}  \color{black}{(b)}}
     \vspace{4.5cm}    
\caption{(a) Dark current in unit 2 Lyman-$\alpha$ channel \textit{versus} temperature. Asterisks show pre-launch measurements, dots show data from calibration campaigns, diamonds show dark current final estimation. (b) An example of a quiet-day Lyman-$\alpha$ signal (level 1). The effect of orbital and daily temperature variations is clearly visible. After subtraction of dark current, the same time-series becomes a flat line. }
   \label{fig:darkcurrent}
\end{figure}

The actual relationship had therefore to be tabulated in smaller steps, exploiting several calibration campaign observations with closed covers, and even some with open covers, when the solar component could be removed. An example is illustrated in Figure \ref{fig:darkcurrent} for the Lyman-$\alpha$ channel of unit 2 (one of the channels that is most affected by temperature effects). A similar relationship was found for each channel. The resulting dark-current estimation is finally subtracted from the data. 

\subsubsection{Correction for Degradation}\label{sec:degradation}
   
\begin{figure}    
   \centerline{\includegraphics[width=0.5\textwidth,angle=0, clip=, trim=5 0 5 0]{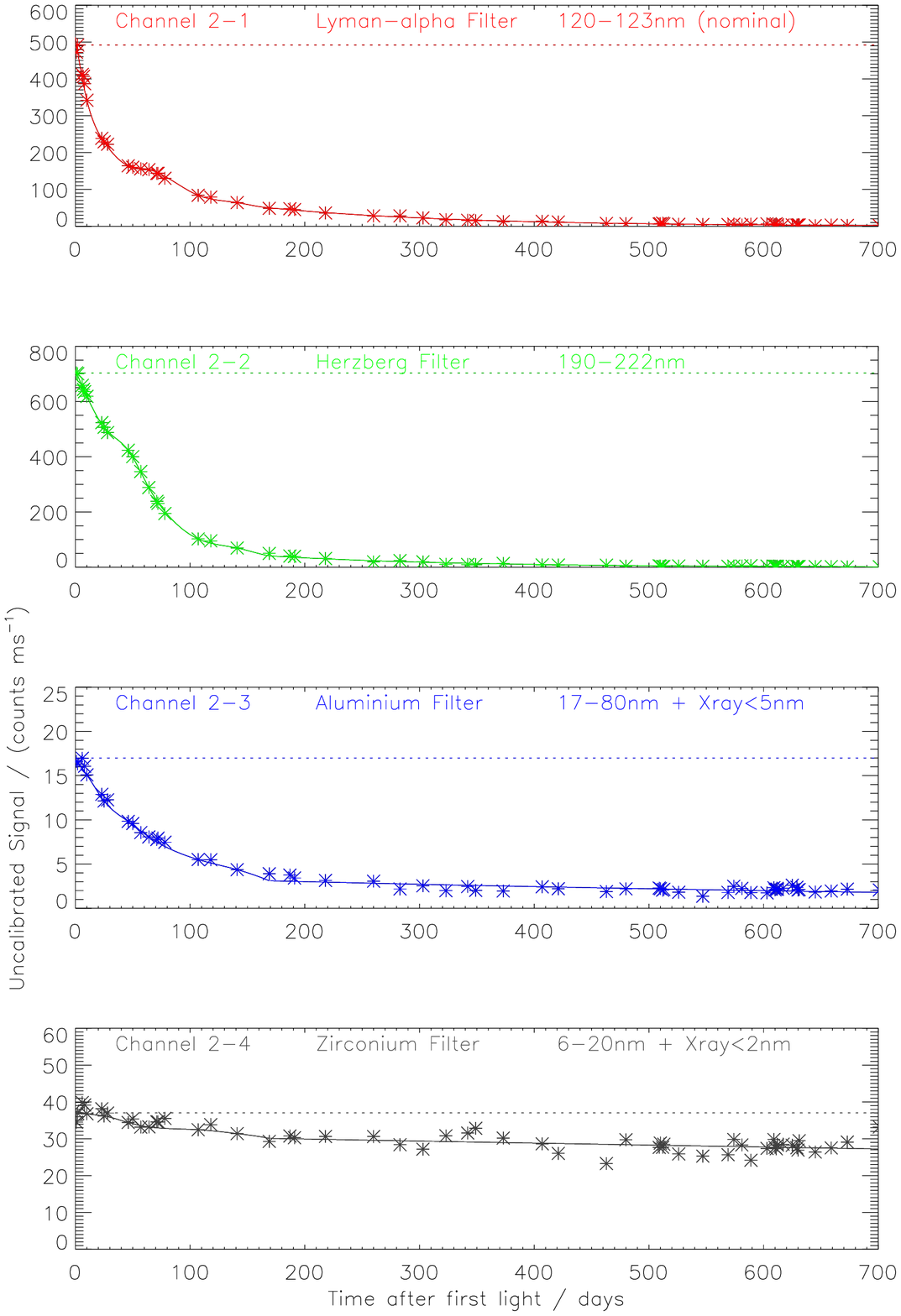} \includegraphics[width=0.5\textwidth,angle=0, clip=, trim= 5 0 5 0]{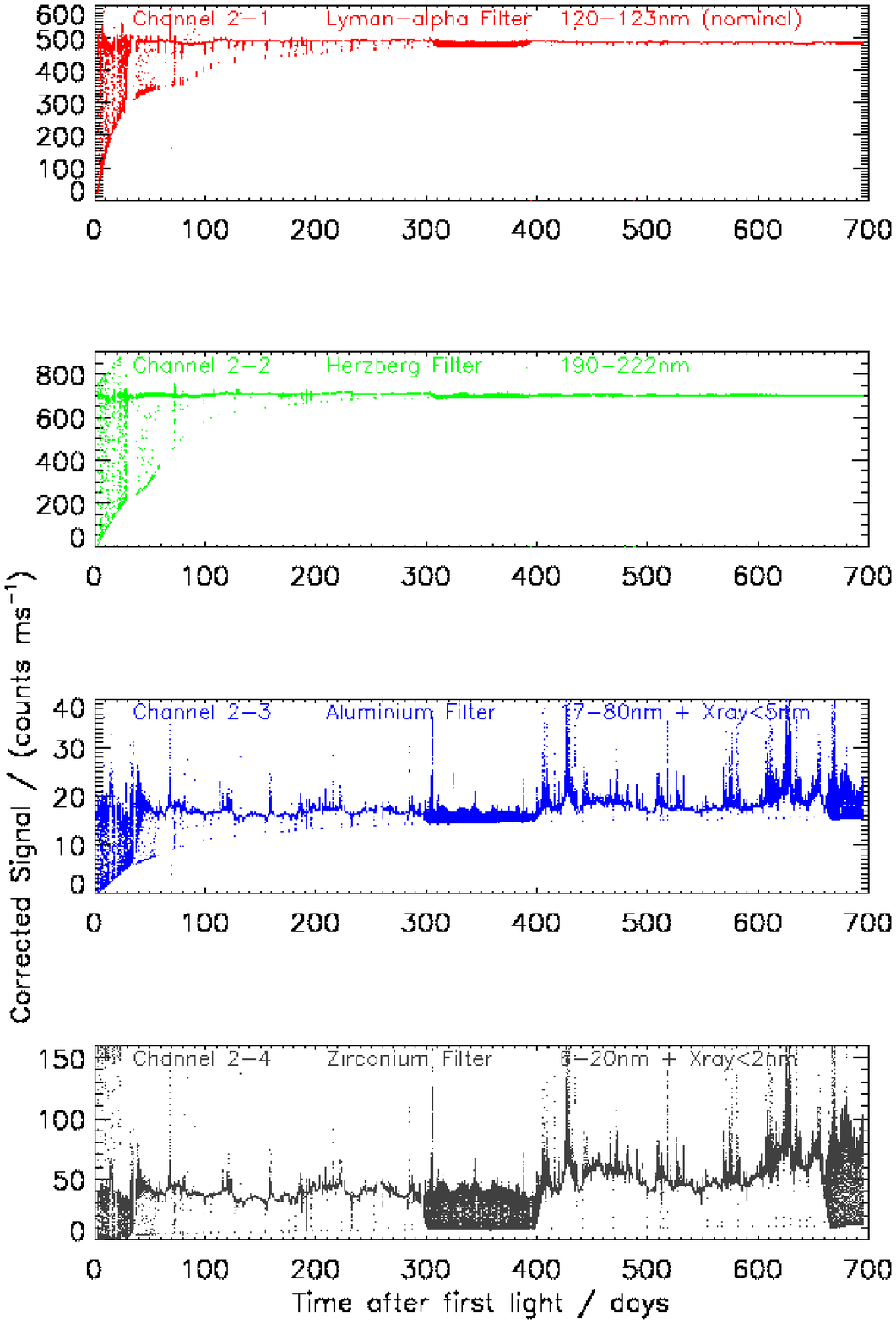}}
\caption{Left column: Degradation of unit 2 estimated from the beginning of the mission. The dark current is subtracted from the signal. Right column: Unit 2 data of the same period, after correction of degradation (and subtraction of dark current). During the season of occultations (days 0 to 40 and 300 to 400), the signal experiences drops to the dark-current level during every orbit. During the whole year, periodic calibration campaigns were scheduled, when the dark current signal was measured. As an unintended side effect of the additive correction for degradation, the level of the dark-current -- which is also the level reached during occultations -- gives the impression of a progressive increase with time, while being actually globally stable.
}
   \label{fig:degradation}
\end{figure}

From the very first hours after the covers were opened, \lyra \- suffered severe degradation, affecting more seriously its longer-wavelength channels (Figure \ref{fig:degradation}). A possible explanation for the degradation observed is condensation and UV-induced polymerization of outgassing molecules on the filter surface. The resulting layer absorbs longer wavelengths more than shorter ones, explaining the different impact on different channels. In the Lyman-$\alpha$ and Herzberg channels of unit 2, the signal fell by 70\,\% within the first month alone and a loss of 99\,\% is now reached, with a degradation process tending to stabilize. The degradation in those channels is now such that solar signal is barely detectable in time series produced by the nominal unit. Fortunately, preserved back-up units are available for observation campaigns. 
It was not possible in the long run to calibrate and compare \lyra \- with other instruments without first taking the degradation into account. This, in turn, meant that the degradation had to be separated from the solar variation.

To some extent, the degradation can be calculated by internal means. For the two shorter-wavelength channels (aluminum and zirconium), this is done with the help of the spare units 1 and 3, which so far were only used for very occasional campaigns. Their covers were only opened for about 50 hours during the whole year 2010, while unit 2 was observing the Sun almost continuously since it saw the first light. The zirconium channels in units 1 and 3 have seen no apparent loss at all at the time of writing and any of them could be used as a reference for both the aluminum and zirconium channels of unit 2, after subtraction of the variations induced by solar activity. The evolution of the aluminum and zirconium channels is correlated.

Unfortunately, there is no such internal reference for the longer-wavelength channels (Lyman-$\alpha$ and Herzberg) of unit 2. Degradation was so rapid in those channels that even units 1 and 3 were affected. The solution that we adopted is to conjecture that the quiet-Sun signal does not show any long-term trend. Any global deviation is considered as an effect of degradation. The data analysis is somewhat biased in this approach, and can only focus on short-term variations of the solar irradiance. 

The evolution of the \lyra \- signal, especially in channels 2-1 and 2-2, shows phases of different degradation velocity: 
\begin{itemize}
\item The degradation trend in the first half year (day 1\,--\,day 169 after first light) is fitted with a spline function through some manually selected data points. There is indeed no apparent mathematical function for the initial degradation; it seems to occur in various phases, and the physical processes behind it (polymerization and emergence of contamination layers on the filters' surfaces) are not well enough known. 
\item The degradation trend for the second half year was at first fitted with a function of the type 1/(a+bt), where $t$ is time and $a$ and $b$ are fitting coefficients. This appeared to be a satisfactory first approach. Other functions, negative exponentials in particular, were also tested. The results of the latter proved to be an even more robust estimate for the future behaviour. A function of the type exp(a+bt) has been introduced in a recent update of the calibration software.
\end{itemize}

After day 100, the degradation has significantly slowed down, however. Therefore, only later data points, representing calibration campaigns after 23 June 2010, are used to estimate the future development. 

The loss caused by degradation is estimated relative to the first light individually for each channel. The degradation is corrected by adding this loss to the current levels. Correction by addition (as opposed to multiplication by a correction coefficient) has advantages, because degradation is a function of the wavelength and does not uniformly affect the broad \lyra \- channels over their whole spectral range. In particular, there is no apparent sensitivity loss in the X-ray range. Flares of similar intensity in \xrs \- scale now peak to the same count rates in aluminum and zirconium channels as they did at the beginning of the mission, while the measurements of the overall EUV background in the same channels has significantly decreased. A multiplicative correction coefficient would artificially exaggerate the flares. On the other hand, there are some disadvantages: the variation of EUV components in channel 3 (and to a lesser degree in channel 4) might be underestimated, and the occultation profiles become distorted since they no longer drop to zero. These effects can be seen in the right panel of Figure \ref{fig:degradation}. 

\subsubsection{Conversion into Physical Units}

Photocurrents measured by \lyra \- detectors can be modeled by 

\begin{equation}\label{eq:rm}
i = i_\mathrm{s} + i_\mathrm{d} = \frac{A}{T} \int_{t} \int_{\lambda} E(\lambda, t) F(\lambda) D(\lambda) \mathrm{d}\lambda \mathrm{d}t + i_\mathrm{d}
\end{equation}

where~:
\begin{itemize}
\item $i$ is the measured photocurrent that can be defined as the sum of a solar $\lbrack i_\mathrm{s}\rbrack$ and a dark current $\lbrack i_\mathrm{d}\rbrack$ contribution
\item $\lambda$ is the wavelength
\item $t$ is the time and is integrated over an exposure
\item $A$ is the aperture area, \textit{i.e.} the area of the detector that is exposed
\item $T$ is the total exposure time (nominally 50\,ms)
\item $E(\lambda, t)$ is the solar spectral irradiance 
\item $ F(\lambda) $ is the filter transmittance
\item $D(\lambda)$ is the detector spectral responsivity
\end{itemize}
and where the integral over $\lambda$ is performed over the whole spectral range in which the instrument is sensitive (not restricted to the defined bandpass), which means that it also includes the out-of-band radiation. This out-of-band radiation constitutes a source of measurement error that might be important in the case of the Lyman-$\alpha$ channel (see purities in Table \ref{tab:channels}).
 
Since \lyra \- channels cover broad spectral ranges, it is not possible to directly invert Equation (\ref{eq:rm}) to retrieve the spectral irradiance $\lbrack E(\lambda, t) \rbrack$ from the measured photocurrents $\lbrack i \rbrack$. For an absolute radiometric calibration, we compare data acquired at any time $\lbrack t \rbrack$ to a pre-degradation reference \lyra \- measurement (the first light data of 6 January 2010), for which this conversion into irradiance units is known. This comparison is performed after correction for the degradation, as detailed in Section \ref{sec:degradation}. It assumes that the relationship between \lyra \- count rate and irradiance in physical units is linear.

\begin{equation}\label{eq:conversionIrradiance}
E_{\mathrm{cal}} = \frac{i_{\mathrm{uncal}} - i_{\mathrm{d}} + corr}{i^{\mathrm{FL}}_{\mathrm{uncal}} - i^{\mathrm{FL}}_\mathrm{d}} E^{\mathrm{FL}}_{\mathrm{cal}}
\end{equation}

where:
\begin{itemize}
\item $E_{\mathrm{cal}}$ and $E^{\mathrm{FL}}_{\mathrm{cal}}$ are the spectral irradiances covering one \lyra \- channel $\lbrack$W m$^{-2}\rbrack$ for respectively the current measurement and the first-light reference 
\item $i_{\mathrm{uncal}}$ and $i^{\mathrm{FL}}_{\mathrm{uncal}}$ are the unprocessed solar irradiance $\lbrack$counts ms$^{-1}\rbrack$ for respectively the current measurement and the first light reference
\item $ i_{\mathrm{d}}$ and $i^{\mathrm{FL}}_\mathrm{d}$ are the dark current measurements $\lbrack$ counts ms$^{-1}\rbrack$ for respectively the current measurement and the first-light reference
\item $corr$ is the corrective term for degradation
\end{itemize}
The conversion of this \lyra \- reference into irradiance units can be done by comparing it to measurements provided by other instruments. 

\begin{figure}  
   \centerline{\includegraphics[width=1\textwidth,clip=]{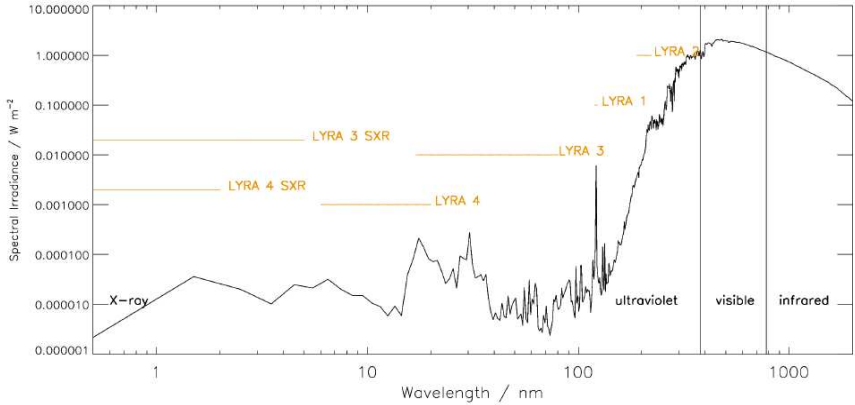}}
   \caption{Reference solar spectrum reconstructed from \see \- and \solstice \- level-3 data on 6 January 2010, used as a reference to infer \lyra \- radiometric calibration. Wavelength coverage of different \lyra \- channels is marked with orange lines.}
   \label{fig:solarspectrum}
\end{figure}

Expressing the first-light reference measurement as physical irradiance units requires knowledge of the detailed solar spectrum for this day. We used a concatenation of spectral irradiance by \see \- from 0.5 nm to 115.5 nm and by \solstice \- from 116.5 nm to 2412.3 nm (see Figure \ref{fig:solarspectrum}).

This spectrum was then inserted in Equation (\ref{eq:rm}) to produce an estimation of the expected \lyra \- photocurrent. The difference of the estimated photocurrent  to the measured one was converted into an excess/default of spectral irradiance with respect to the one observed by SEE and SOLSTICE (integrated over the \lyra \- spectral range). 

In other words, we calculated the \lyra \- first-light spectral irradiance in a given channel using

\begin{equation}\label{eq:firstlight}
E^{\mathrm{FL}}_{\mathrm{cal}} = \frac{i^{\mathrm{FL}}_{\mathrm{uncal}} - i^{\mathrm{FL}}_\mathrm{d}}{\frac{A}{T} \int_{t} \int_{\mathrm{sol.spec.}} E_{\mathrm{S}}(\lambda, t) F(\lambda) D(\lambda) \mathrm{d}\lambda \mathrm{d}t}  \int_{\mathrm{bandpass}} E_{\mathrm{S}}(\lambda) \mathrm{d}\lambda
\end{equation}
where $E_{\mathrm{S}}$ is the solar spectral irradiance (full spectral resolution) from  SEE-SOLSTICE on 6 January 2010.

In Equation (\ref{eq:firstlight}), the integrals over $\lambda$ cover either the whole solar spectrum (denominator) or only the official spectral range of a channel (numerator), since the latter estimates what would have been measured by a perfect instrument.

An additional complication came from the fact that, in the case of the Lyman-$\alpha$ channel, the excess values computed for the three redundant units differed significantly (unit 1: +81.3 \,\%, unit 2: +91.2\,\%, unit 3: +3.3\,\%). Moreover, it is difficult to compare units 1 and 2 (MSM diamond detectors) to unit 3 (Si detector), since the diamond detectors have an additional second peak around 200\,nm, while the silicon detector collects 70\,\% of its non-nominal input between 200 and 1100\,nm, with a peak between 900 and 1000\,nm. Therefore, it is hard to make a statement such as ``\lyra \- observes $x$\,\% more irradiance as compared to \solstice'' -- which explains the question mark in Table \ref{tab:excessirradiance}. For this channel, the \solstice \- value for the nominal interval 120\,--\,123 nm must be assumed. 

The other channels showed more consistency, so we used the average value over the three units (see Table \ref{tab:excessirradiance})

\begin{table}
\caption{Excess of spectral irradiance as observed by \lyra \- in comparison to \see \- and \solstice \- measurements.
}
\label{tab:excessirradiance}
\begin{tabular}{c|c|c|c} 
 Lyman-$\alpha$ & Herzberg & Aluminum & Zirconium \\
 \hline
 ?  & +18.0 \,\% & +13.3\,\% & +9.2\,\% \\
\end{tabular}
\end{table}

Combining Equations (\ref{eq:conversionIrradiance}) and (\ref{eq:firstlight}), and taking into account that the solar irradiance might be considered as constant over sub-second periods, we obtain:
\begin{equation}\label{eq:fullmeasure}
E_{\mathrm{cal}} = \frac{i_{\mathrm{uncal}} - i_\mathrm{d}+corr}{A \int_{\mathrm{sol.spec.}} E_{\mathrm{S}}(\lambda) F(\lambda) D(\lambda) \mathrm{d}\lambda} \int_{\mathrm{bandpass}} E_{\mathrm{S}}(\lambda) \mathrm{d}\lambda
\end{equation}

\subsection{Non-Solar Features in the \lyra \- Data}\label{s:features}
Undesired features, such as imprints of large angle rotations of the spacecraft or perturbations due to the South Atlantic Anomaly (SAA), are present in \lyra \- data and must not be interpreted as solar signal variation.  This section provides a list of those features.

\subsubsection{Flat-Field Effects}

As described in Section \ref{s:calibration}, the calibration process does not yet correct the data for variations associated with pointing fluctuations. An analysis of \proba \- attitude over several orbits reveals that pointing is stable up to 90 arcseconds. Spacecraft jitter introduces fluctuations in the \lyra \- signal of less than 1\,\%.

\begin{figure}[h!]
   \centerline{\includegraphics[width=1.0\textwidth,clip=]{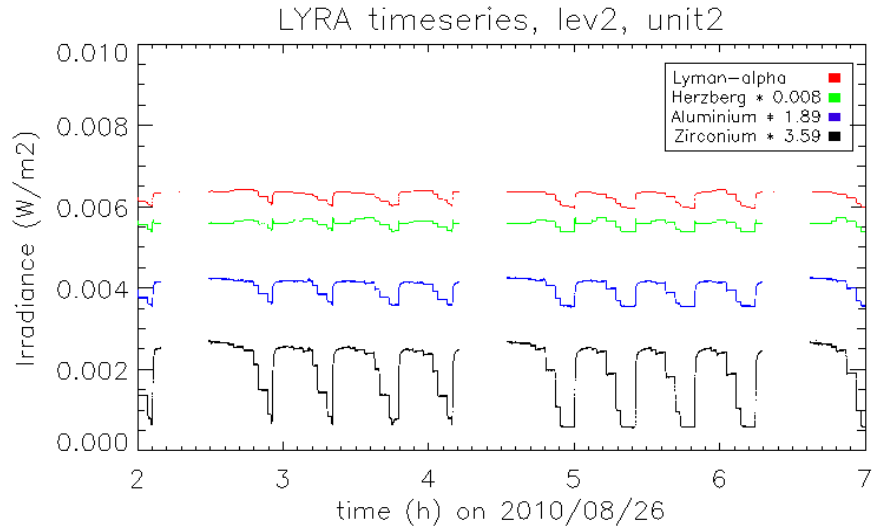}
              }
   \caption{Flat-field campaign of 26 August 2010: the spacecraft is off-pointed from 0$^\circ$ to 3$^\circ$, in steps of 0.5$^\circ$, successively in the S, E, N, W, SE, NE, NW, and SW directions. Results are plotted for unit 2 channels. Time series have been rescaled to fit the same range and appear in the same order as in the legend. Scaling coefficients are indicated in the legend.
               }
   \label{fig:ff}
\end{figure}

Nevertheless, it often happens that \proba \- is off-pointed in the framework of calibration or scientific campaigns, introducing signal fluctuations of which the amplitude depends on the new pointing (see Figure \ref{fig:ff}). Such campaigns are noted in the public calendar of \lyra \- activities linked from the website of the instrument \url{http://proba2.oma.be/about/operations}. It is intended to correct for those fluctuations in a future version of the calibration routine. 

\subsubsection{Large-Angle Rotations of the Spacecraft}
Four times per orbit, the spacecraft rotates 90$^\circ$ around the axis pointing toward the Sun, to avoid the Earth shadowing its star trackers. Because of the inhomogeneous flat field of the diamond detectors, such rotations are clearly visible in the time series (see Figures \ref{fig:occultationfeature} to \ref{fig:aurora}). Unfortunately, the acquisition of pointing parameters is performed at a limited cadence, not fast enough to deduce the spacecraft movement during such maneuvers with a sufficient accuracy and to allow for any flat-field correction. Large-angle rotations are systematic and cannot therefore be confused with natural solar variability.
In the future, we are considering removing those features from the data.

\subsubsection{Occultations}

\begin{figure}    
   \centerline{\includegraphics[width=1.0\textwidth,angle=0, clip=, trim=0 0 0 0]{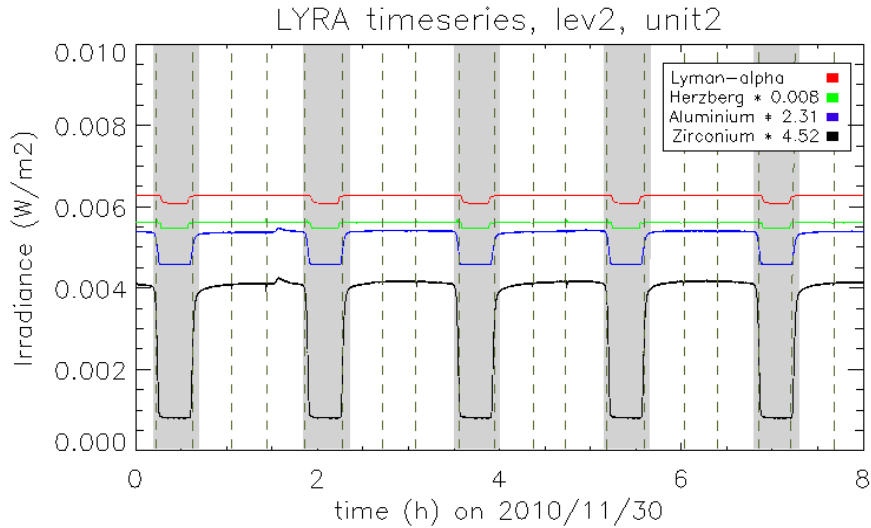} 
   }
   \caption{Drop of signal when the spacecraft transits the Earth's shadow (highlighted with gray). Vertical dashed lines indicate the large-angle rotations of the spacecraft. Time series have been rescaled to fit the same range and appear in the same order as in the legend. Scaling coefficients are indicated in the legend.}
   \label{fig:occultationfeature}
\end{figure}

From November to February (approximately), the orbit of \proba \- crosses the Earth's shadow. This produces progressive attenuation of the solar signal when \lyra \- is observing the Sun through deeper layers of the Earth's atmosphere and ends up with a total extinction, see Figure \ref{fig:occultationfeature}. As for large-angle rotations, those features are easily identifiable by an observer because of their regularity.

\subsubsection{Slow Stabilization of MSM Detectors}
MSM detectors need time to stabilize when exposed to light (see, \textit{e.g.}, Figure \ref{fig:SAA}). We can explain this phenomenon by the existence of surface defects, which trap the photoelectrons produced, preventing them from being collected at the electrodes. The same happens when closing the covers: the MSM detector signal does not drop immediately to zero. Trapped electrons may even take several hours to leak out.   

\subsubsection{Perturbations in the \lyra \- Electronics}
\begin{figure}    
   \centerline{\includegraphics[width=1.0\textwidth,angle=0, clip=, trim=0 0 0 0]{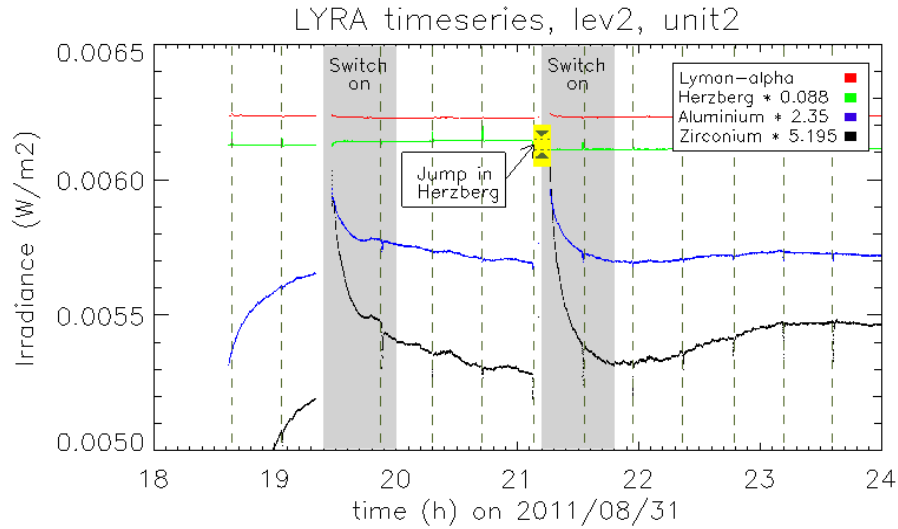}
   }
\caption{Typical perturbation in instrument electronics. The peaks in irradiance measurements in the aluminum and zirconium channels (highlighted with gray) are caused by switching the instrument on. A small jump in Herzberg channel irradiance is also visible. The occurrence of such offsets is synchronous with a change of acquisition parameters (\textit{e.g.} change of the unit used, of the acquisition cadence or of the cover status). Dashed vertical lines indicate the large-angle rotations of the spacecraft. Time series have been rescaled to fit the same range and appear in the same order as in the legend. Scaling coefficients are indicated in the legend.} 
   \label{fig:switch}
\end{figure}

Switching on the instrument can produce a peak in the signal measured in some channels. This effect is likely to happen any time that we change the operational mode, \textit{i.e.} when activating a new unit, performing a calibration campaign, reloading the onboard Field-Programmable Gate Arrays (FPGA) or coming back to nominal acquisition after one of the aforementioned activities. Such peaks are illustrated in Figure \ref{fig:switch}. 

\subsubsection{Jumps of Irradiance Observed in the Herzberg Channel}
From time to time, the signal measured in the nominal Herzberg channel jumps from one level to another (small offset added or removed, see Figure \ref{fig:switch}). The reason for these jumps is not completely understood at the moment, but it is synchronous with the release of onboard commands. Jumps definitely complicate the exploitation of nominal Herzberg time series, whenever a high accuracy is required for solar irradiance.

\subsubsection{South Atlantic Anomaly Perturbations}
When transiting the South Atlantic Anomaly (SAA), secondary electrons generated by high-energy protons hit the detectors and make \lyra \- signal more noisy. This effect mostly affects Si detectors, independently of the spectral range, while MSM and PIN detectors -- being radiation hard -- usually do not show significant perturbations (the effect of SAA on \lyra \- time series is visible in Figure \ref{fig:SAA}). It is worth mentioning here that four channels of \lyra \- are amplified tenfold on-board to make up for their otherwise low signal: all three Lyman-$\alpha$ channels and the zirconium channel of unit 2. In these channels, the SAA perturbations appear magnified.

\subsubsection{Auroral Perturbations}
\begin{figure}    
   \centerline{\includegraphics[width=1.0\textwidth,angle=0, clip=, trim=0 0 0 0]{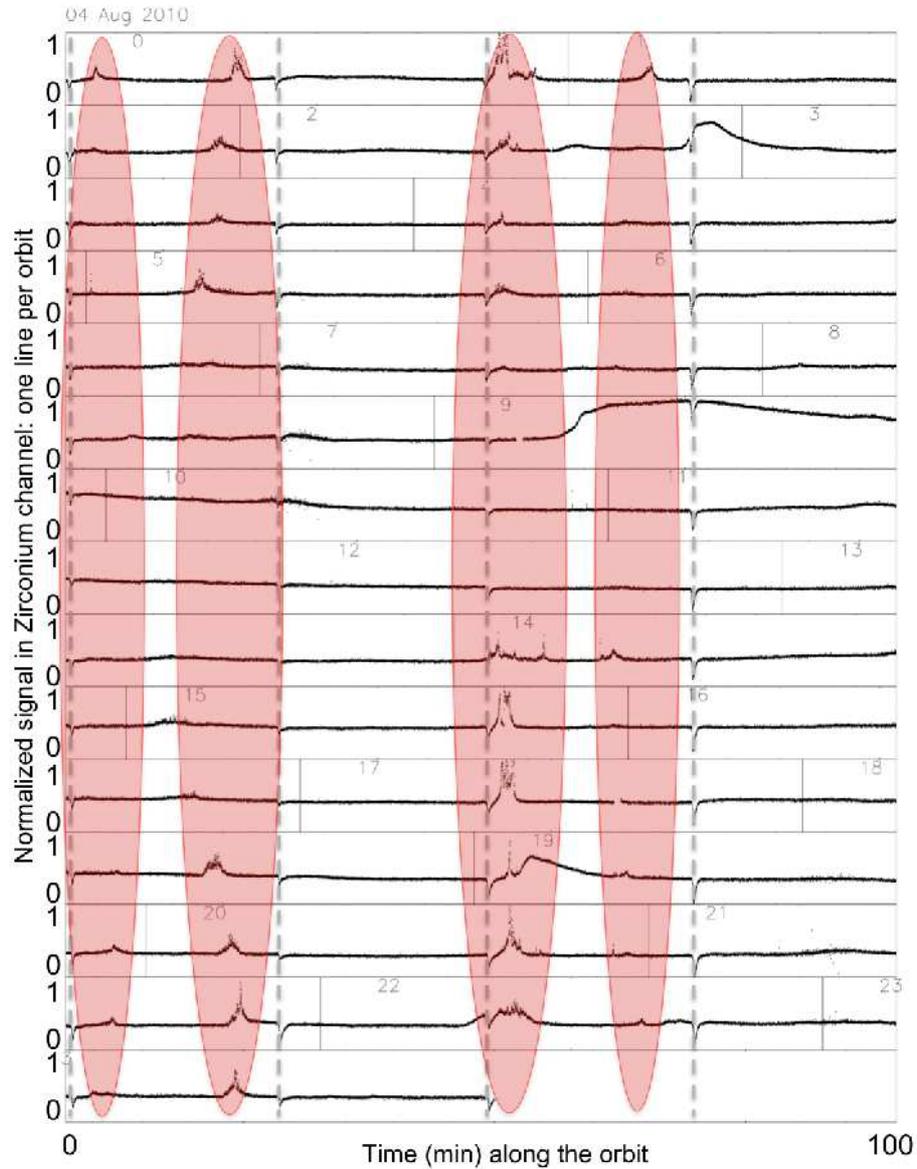}
   }
\caption{Perturbations appearing when crossing the auroral zone during or after a geomagnetic storm. Each horizontal line corresponds to one orbit. Four zones of perturbations are identified for each orbit (highlighted with red). They occur systematically in the same range of latitudes (north and south auroral ovals are each crossed twice during an orbit). Dashed vertical lines indicate the large-angle rotations of the spacecraft. } 
   \label{fig:aurora}
\end{figure}

Auroral zones usually do not have any impact on the \lyra \- signal, with one noticeable exception: such perturbations appear in case of geomagnetic-storm conditions, usually when the Kp index is above four (see Figure \ref{fig:aurora}). We have not yet identified what mechanism provokes the perturbations that we observe in \lyra \- data. Are they caused by auroral particles, subsequent bremsstrahlung, or auroral photons? The question is still under investigation. It is quite remarkable that it affects equally diamond and silicon detectors, but not equally all spectral ranges: aluminum and zirconium channels are affected, Lyman-$\alpha$ and Herzberg are not. On the other hand, the imager \swap \- onboard \proba \- \cite{2012SoPh..tmp..217S}, although observing at 17.4 nm, does not exhibit this effect; this may be related to the fact that \lyra, unlike \textsc{Swap}, has a direct optical path, without any reflection by mirrors. In \swap, the mirrors inserted in the optical path might absorb energetic particles, preventing them to reach the detector. 

\subsection{Radiometric Accuracy}

From Equation (\ref{eq:fullmeasure}), the maximal uncertainty on the calibrated LYRA data is described by 
\begin{equation}\label{eq:accuracy}
\frac{\Delta E_{\mathrm{cal}}}{E_{\mathrm{cal}}} = \frac{\Delta i}{i} + \frac{\Delta \left ( \int_\lambda E \mathrm{d}\lambda \right )}{\left ( \int_\lambda E \mathrm{d}\lambda \right )} + \frac{\Delta A}{A} + \frac{\Delta \left ( \int_\lambda E F D \mathrm{d}\lambda \right )}{\left ( \int_\lambda E F D \mathrm{d}\lambda \right )} 
\end{equation}
where $i = i_{\mathrm{uncal}} + i_{\mathrm{d}} + corr$. 
We will discuss the terms of this equation one by one.
\subsubsection{Error on the Measured Currents}
The first term of Equation (\ref{eq:accuracy}) takes into account the measurement error on the solar signal and on the dark current, as well as an estimation of the error introduced by the corrective term. Both solar signal and dark current depend on:
\begin{itemize}
\item a quantization error
\item the stability of three onboard reference voltages (0\,V, 2.5\,V and 5\,V), which are used to convert data expressed in count number to voltages. This conversion implies a polynomial fit (second order) of the three reference voltages, which also introduces its own error
\item the uncertainty on the internal resistor, used to convert the voltages into currents
\item the uncertainty on the integration time, which is related to the quartz stability
\end{itemize}
\inlinecite{2009A&A...508.1085B} estimated the relative uncertainty associated with these four parameters for each LYRA channel, and obtained an associated error of $0.03 \,\%$ at the maximum.
\newline
Additionally, $i_{\mathrm{uncal}}$ is also affected by the pointing stability (jitter). For Sun-centered acquisitions, the jitter combined with flat-field non-homogeneities result in an uncertainty $\le 1\,\% $ in all channels. 
\newline 
Furthermore, $i_{\mathrm{uncal}}$ can be split into its in-band and out-of-band components. The latter is considered as part of the measurement error and is of the same order as the complement to the purity in Table \ref{tab:channels}. This is one of the major sources of uncertainty since it is of the order of 5\,\% of $i_{\mathrm{uncal}}$ in most channels and even of 75\,\% in Lyman-$\alpha$.
\newline
Last but not least, the error associated to the corrective term is hard to estimate. But this term is based, for the aluminum and zirconium channels, on the assumption that the unit 3 zirconium channel did not degrade, and for Lyman-$\alpha$ and Herzberg channels on the assumption that they do not show any long-term trend. Therefore, one might consider that the error introduced by the corrective term is of the order of the actual degradation of the unit 3 zirconium channel (a few per cent of the corrective term) for aluminum and zirconium channels, and of the maximum variation over the solar cycle for Lyman-$\alpha$ (between 40 and 100 \,\% of the corrective term) and Herzberg (about 5\,\%). 
\newline
In conclusion, the uncertainty on the measured current is mostly due to the correction for degradation and to the out-of-band radiation, with the importance of the first one progressively increasing over the mission, as the second one decreases. On average, we consider an overall uncertainty on the measured current of about 100\,\% in Lyman-$\alpha$ and of 6\,\% in the other channels.
\subsubsection{Error on the Spectrum}
The second term in Equation (\ref{eq:accuracy}) depends on the accuracy of \solstice \- and \see \- measurements, which are of the order of 5\,\% and 10\,--\,20\,\% respectively. The first value is applicable to Lyman-$\alpha$ and Herzberg channels, while the second one is used for aluminum and zirconium channels.
\subsubsection{Error on the Aperture Area}
The aperture area was measured at the Swiss Federal Office of Metrology and Accreditation (METAS), with an uncertainty of 0.07\,\% (see  \inlinecite{2009A&A...508.1085B} ).
\subsubsection{Error on the Simulated Current}
Combining the uncertainty on the filter and detector characterization, which is provided in  \inlinecite{2009A&A...508.1085B}, with that of the solar spectrum results in an estimated error of 15\,--\,20\,\% in all channels
\subsubsection{Conclusion}
All in all, we consider the uncertainty on \lyra \- calibrated data to be of the order of 30\,--\,40\,\% for all channels except Lyman-$\alpha$, where it is about 120\,\%.

\begin{figure}[h!]
   \centerline{\includegraphics[width=1.0\textwidth,clip=]{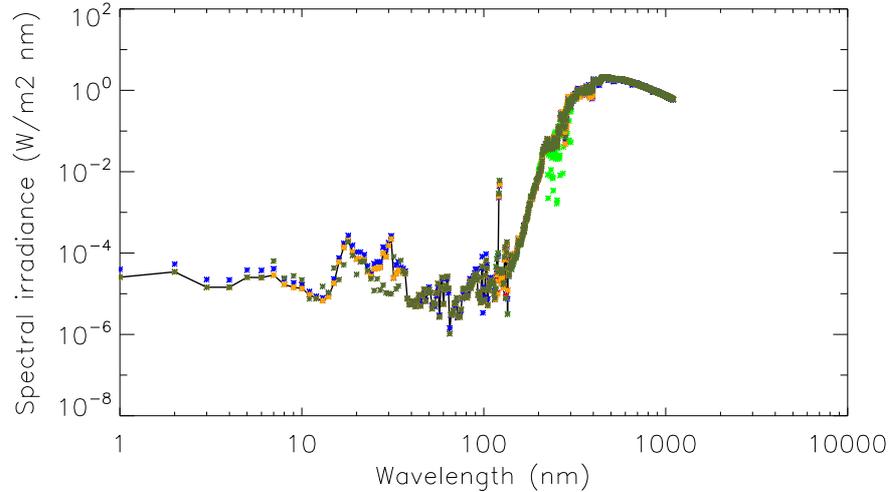}
    }
   \caption{Alternative spectra used to calibrate \lyra \- data. The full black line represents the \see-\solstice \- reference spectrum. Alternative spectra that were tested are over-plotted and respectively correspond to NRLSSI (blue), SRPM (green), SATIRE (orange), and EVE (olive).
               }
   \label{fig:spectraFL}
\end{figure}

In an attempt to validate those uncertainties, we have used alternative spectra in Equation (\ref{eq:fullmeasure}) and have seen how \lyra \- data were affected. Unfortunately, for the considered date (6 January 2010) and spectral ranges, we are not aware of any other measured spectrum than the \see  \- and \solstice \- ones. Even empirical models such as NRLSSI (\opencite{2005SoPh..230...27L}) and SATIRE (\opencite{2009JGRD..11400I04K}) are not provided for periods after 2006 and 2007 respectively. We have therefore tried to find a date with a similar solar-activity context as during \lyra \- first-light, for which those models produced spectra. We chose 10 November 2005. We also tested the SRPM model (\opencite{2009ApJ...707..482F}), which applies for very quiet-Sun conditions. Additionally, to also cover the shortest wavelength ranges in \lyra,  we picked a date after 6 January 2010 when \eve \- was in use and repeated the exercise. The selected day is 16 June 2010. The obtained spectra are plotted in Figure \ref{fig:spectraFL}. 
While NRLSSI affected the four channels of \lyra, SATIRE and SRPM only concerned the wavelengths above 115 and 200\,nm and therefore had a limited impact on the aluminum and zirconium channels. \eve \- (6\,--\,36\,nm) on the other hand only had an impact on these two channels. The variations induced in \lyra \- measurements are summarized in Table \ref{tab:impactSpectrum} and are all within the cumulative error related to spectrum and simulated current.
\begin{table}
\caption{Variation percentage in \lyra \- calibrated data induced by the use of an alternative reference spectrum (these values are for \lyra \- nominal unit).
}
\label{tab:impactSpectrum}
\begin{tabular}{c c c c c} 
 Spectrum & Lyman-$\alpha$ & Herzberg & Aluminum & Zirconium \\
  NRLSSI & -17.5\,\% & -0.4\,\% & -7.8\,\% & +0.6\,\% \\
  SATIRE & +19.1\,\% & -3.5\,\% & 0\,\% & 0\,\% \\
  SRPM & +2.3\,\% & -4.2\,\% & 0\,\% & 0\,\% \\
  EVE & - & - & +3.7\,\% & +24.1\,\% \\
\end{tabular}
\end{table}

\section{Science Opportunities}\label{s:sc_opportunities}

In this section, we briefly describe the main topics of the scientific exploitation of the \lyra \- data.

\subsection{Flares}

\begin{figure}[h!]  
   \centerline{\includegraphics[width=0.57\textwidth,angle=0, trim=0 10 0 10]{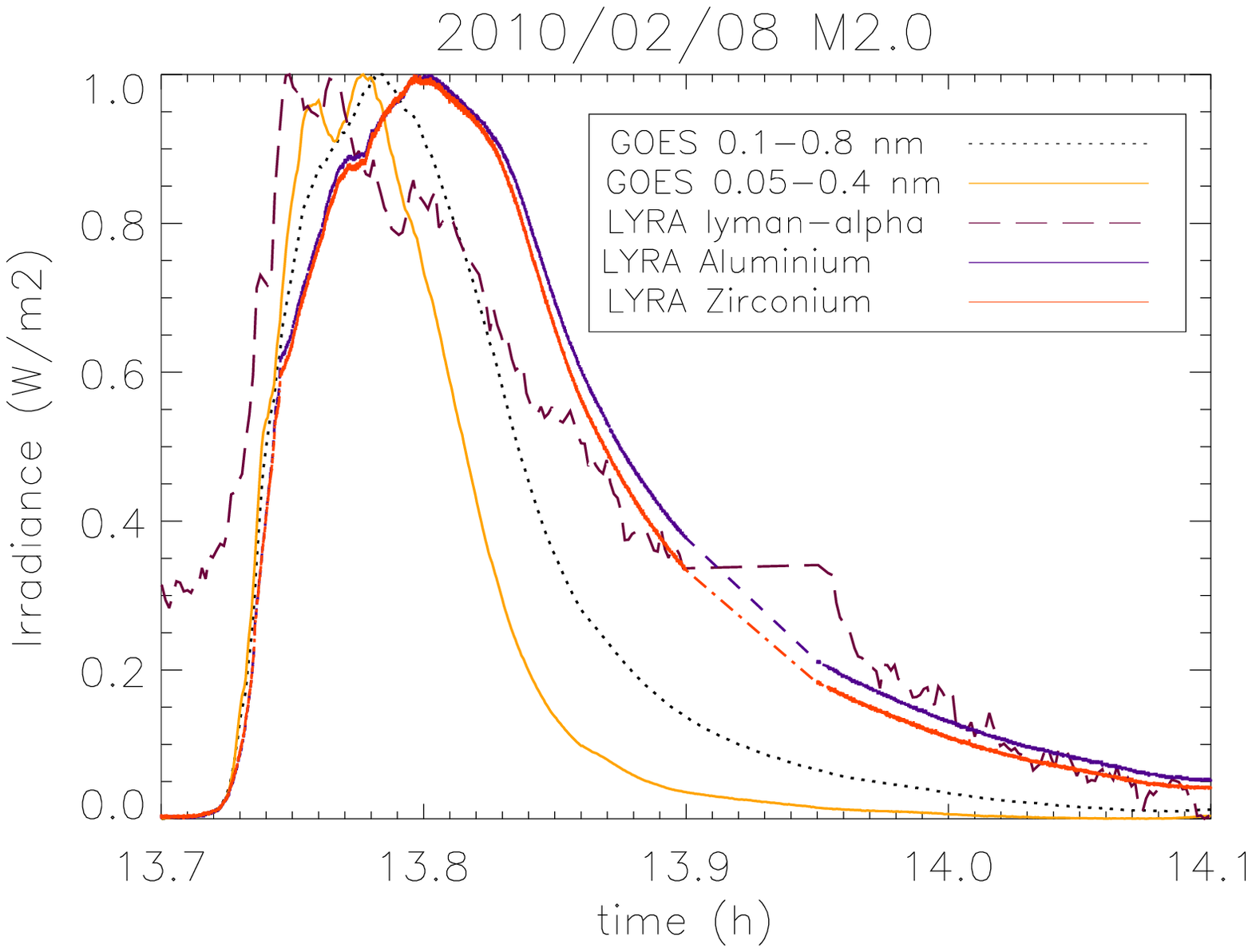} \vspace{-0.3cm}\hspace{-0.3cm}\includegraphics[width=0.52\textwidth,angle=0, trim=0 10 0 10]{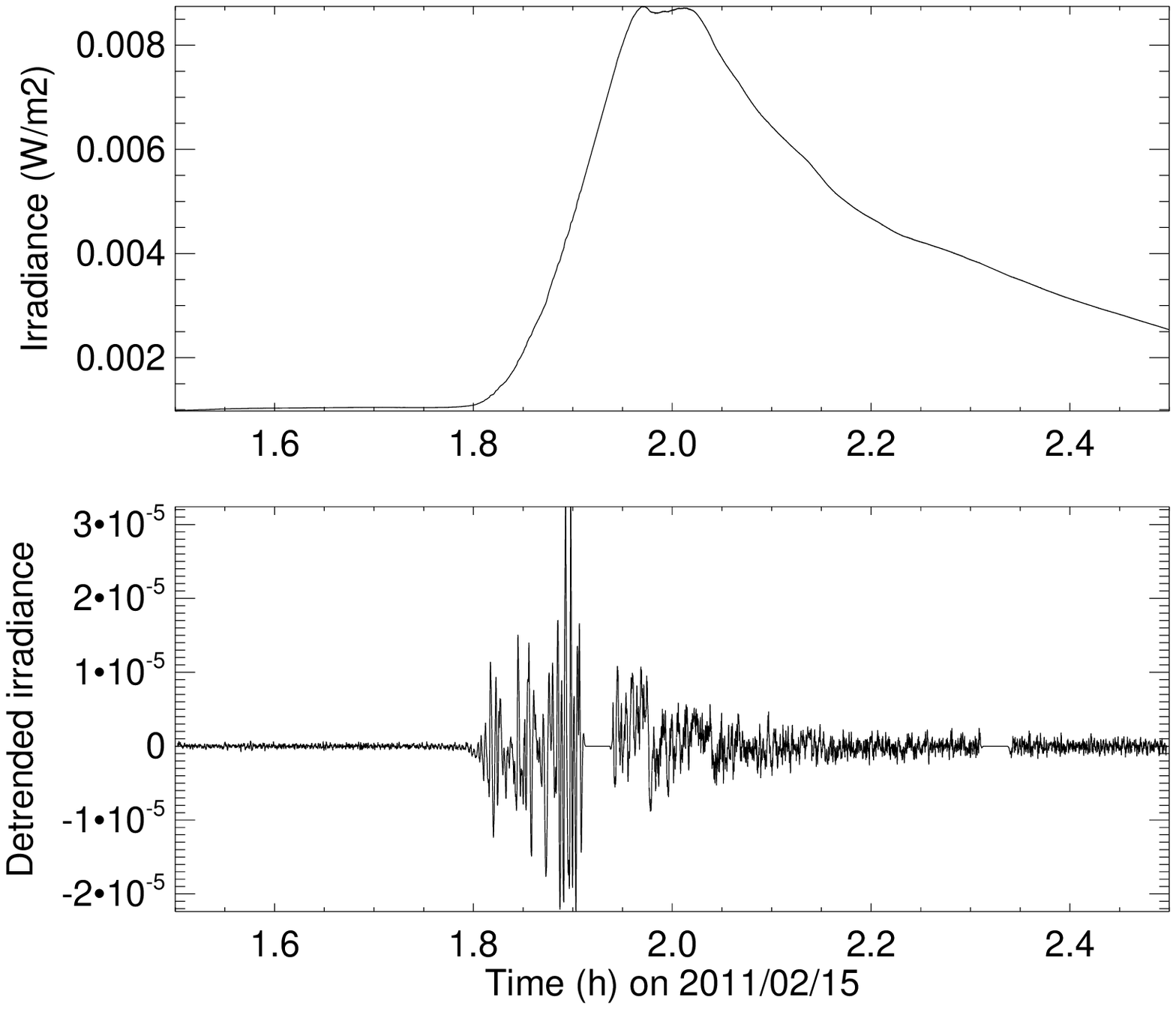}}
     \vspace{-4cm}   
     \centerline{ \bf     
      \hspace{0.06\textwidth}   \color{black}{(a)}
      \hspace{0.48\textwidth}  \color{black}{(b)}
         \hfill}
     \vspace{4cm}    
\caption{Flare analysis with \lyra \- data: (a) shows the differences between start-, end-, and peak-time of a flare when looking at various spectral ranges, from soft X-rays (\xrs) to EUV (\lyra). For clarity, data corresponding to a large angle rotation of the \proba \- spacecraft between 13:54 and 13:57 have been removed. (b) highlights oscillations detected during the onset of the X2.2 flare of 15 February 2011. The upper panel shows the flare curve in the \lyra \- zirconium time series, while the lower panel shows the same observations detrended by subtracting the signal smoothed using a 20-second boxcar. }
   \label{fig:flares}
\end{figure}

Flares are one of the main scientific targets of \lyra. Comparing the flare profiles in \lyra \- channels with observations of instruments acquiring at other wavelengths \xrs, \eve $\ldots$) allows one to determine a chronology of the temperature evolution in the flare and to confront it with the theoretical scenarios for energy release (see Figure \ref{fig:flares}, panel (a)). 

In addition, the \lyra \- high sampling rate allows one to detect short-timescale phenomena, such as quasi-periodic pulsations (QPP) (Figure \ref{fig:flares}, panel (b)) -- with periods from fractions of a second to a few minutes -- that are observed in flares in many wavelengths (see \opencite{2011ApJ...740...90V}; \opencite{ApJ_Oscillations_Dolla}). Here again, \lyra \-  can help with completing the picture when trying to understand the origin of these QPP.

\subsection{Variability of Solar Irradiance}

The Sun's primary influence on Earth is through its radiation, and monitoring and understanding its variations are of prime importance. It is now well accepted that monitoring the solar spectral irradiance over the long-term is necessary in order to understand the role of the Sun in Earth's climate change, especially the EUV part of solar spectrum, which strongly affects the status of Earth's ionosphere. Such long-term monitoring exceeds the capabilities of one single space instrument. It can only be achieved by combining observations of different missions. Together with \euvs, \sem, \see, \solstice, and \eve, \lyra \- is part of the picture. In addition, most of its channels are interesting candidates to serve as proxies for reconstructing the whole solar irradiance in the EUV (see \opencite{2009GeoRL..3610107D} and \opencite{2011A&A...528A..68C}). 

\subsection{Sun--Moon Eclipses}

\lyra \- measurements of irradiance profiles during solar eclipses are a matter of special interest. The irradiance data acquired during solar eclipses allow us to assess the center-to-limb variations (CLV) of the solar brightness. As the radiation from different disk positions originates at different heights, the CLV curves allow to sample the broad range of layers of the solar atmosphere. This provides useful information for testing and refining solar atmosphere models (\textit{cf.}  \opencite{2005SoPh..229...13N}; \opencite{2008ApJ...680..764K}).

\begin{figure}[h!]
   \centerline{\includegraphics[width=1.0\textwidth,clip=]{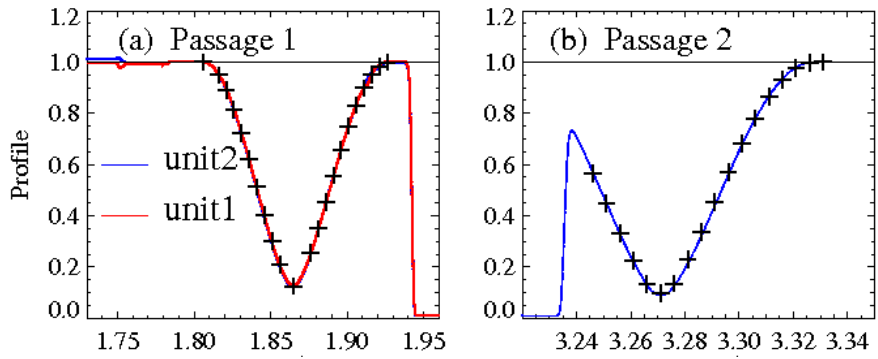}
    }
   \caption{Relative irradiance data of Herzberg channel during the annular eclipse on 15 January 2010. Time is in hours after 15 January 2010 00:00 UT. (a) Passage 1 observed by unit 1 and unit 2. (b) Passage 2 observed only by unit 2. The crosses indicate the profiles calculated with the \textsf{COSI} code.
               }
   \label{fig:eclipse}
\end{figure}

Up to mid-2012, \lyra \- observed seven solar eclipses: two in 2010 (15 January and 11 July), four in 2011 (4 January, 1 June, 1 July, and 25 November), and one in 2012 (20 May). The 15 January 2010 eclipse was the longest annular solar eclipse of this millennium. The radiative transfer COde for Solar Irradiance (\textsf{COSI}: \opencite{2008A&A...492..833H} and \opencite{2010A&A...517A..48S}) was used to model the CLV of the solar radiation measured in the \lyra \- Herzberg channel and to model the light curves of the 15 January 2010 eclipse. The comparison of measured and calculated light curves is presented in Figure \ref{fig:eclipse}. One can see that the calculations with \textsf{COSI} are in excellent agreement with the \lyra \- measurements; see \inlinecite{TI_PROBA2_<Eclipses>_<Shapiro>} for a more complete analysis of eclipses.

\subsection{Occultations}\label{s:occultations}

\begin{figure} [h!]
   \centerline{\includegraphics[width=0.5\textwidth,angle=0, clip=, trim= 0 0 10 0]{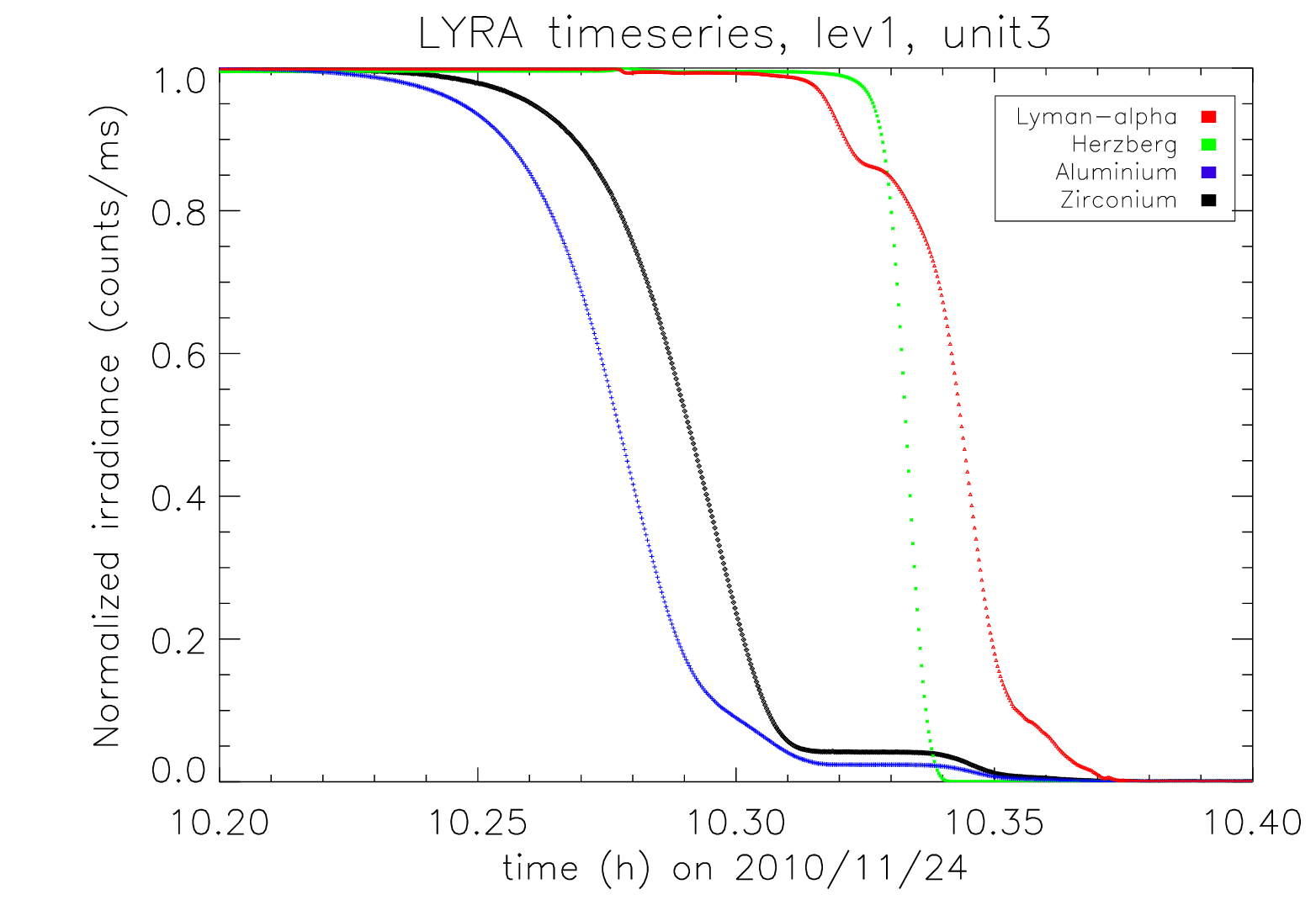}\includegraphics[width=0.57\textwidth,angle=0, clip=, trim=20 20 20 0]{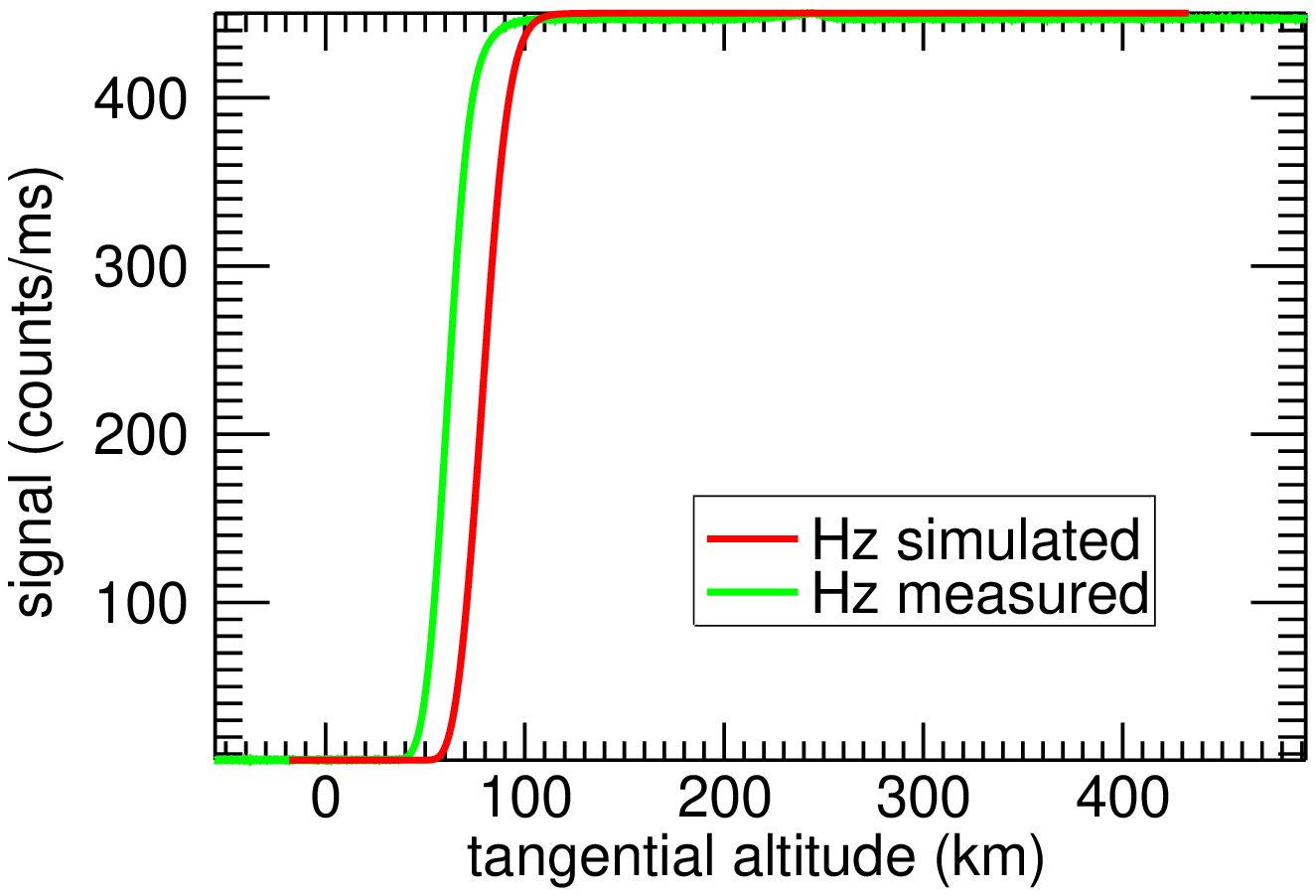}}
     \vspace{-4cm}   
     \centerline{ \bf     
      \hspace{0.5cm}   \color{black}{(a)}
      \hspace{0.45\textwidth}  \color{black}{(b)}
         \hfill}
     \vspace{3.7cm}    
\caption{Panel (a) shows the normalized extinction curves when \lyra \- is progressively observing the Sun through deeper layers of the Earth's atmosphere (\textit{i.e.} when the tangential altitude is decreasing), for all channels. Panel (b) shows comparison between modeled and measured Herzberg-channel extinction curve for increasing tangential altitudes. 
}
   \label{fig:occultation}
\end{figure}

As the Sun approaches the Earth's shadow, \lyra \-'s line-of-sight crosses deeper layers of the Earth's atmosphere. There, atmospheric constituents absorb part of the solar radiation in a range of wavelengths that is intrinsically dependent of their chemical nature (Figure \ref{fig:occultation} (a)). Analyzing the extinction curve of each \lyra \- channel therefore provides information about the distribution of some of theses constituents as a function of altitude. The main molecules active in \lyra \- bandpasses are O, O$_2$, O$_3$, N$_2$, and H$_2$O. In Figure \ref{fig:occultation} (b), we provide an example of the reconstruction of Herzberg-signal extinction by O$_2$ and O$_3$ (see \inlinecite{Dominique_OPAC3_2008}, and references therein). 

Observing the extinction curves also allows one to detect the contamination of a spectral channel by longer wavelengths. Monitoring the evolution of such a curve therefore provides a reliable way to track any spectral degradation of the channel and hopefully to correct for it.

\subsection{Space Weather Monitor}
The solar EUV flux informs us about the occurrence of solar events and forms a vital input for aeronomy models \cite{2005AnGeo..23.3149H}. Consequences of such events might affect, among others, telecommunication or spacecraft trajectories \cite{2007swpe.book.....B}. With its quasi-continuous visibility of the Sun (the above-mentioned short occultation periods excepted), its high-cadence and its approximately nine contacts with the ground per day, \lyra \- constitutes a valuable space-weather monitor.

\section{Conclusion}
After more than two years of operations, \lyra \- demonstrated its ability to produce time series of solar irradiance in the XUV--EUV--MUV range with a very favorable sampling rate. Uncalibrated and calibrated data products, as well as various quicklooks and a list of flares, are distributed via the instrument website (\url{http://proba2.sidc.be}). Among others, the instrument appears to be particularly well adapted to the analysis of flares, for which its spectral channels and sampling rate are ideal. High sampling rate also benefits other fields of research, of which some have been mentioned in Section \ref{s:sc_opportunities} (analysis of the center-to-limb brightness variation, determining the composition of Earth's upper atmosphere). Finally, \lyra \- finds its place in the long-term monitoring of solar irradiance in the EUV--UV range.

%
 \begin{acks}
The authors acknowledge the support from the Belgian Federal Science Policy Office through the ESA-PRODEX programme. \lyra \- is a project of the Centre Spatial de Li\`ege, the Physikalisch-Meteorologisches Observatorium Davos, and the Royal Observatory of Belgium, funded by the
Belgian Federal Science Policy Offce (BELSPO) and by the Swiss Bundesamt
f\"ur Bildung und Wissenschaft. The authors want to acknowledge this consortium, as well as the whole \lyra \- team for their investment and motivation. A special thanks goes to D. McMullin and J. Zender for their pertinent suggestions.
 \end{acks}

%
%
%

\end{article} 
\end{document}